\documentclass[aps,twocolumn,letterpaper]{revtex4-1}

\usepackage{amsmath}
\usepackage{amssymb}
\usepackage{amsfonts}
\usepackage{graphicx}
\usepackage{dcolumn}
\usepackage{float}
\usepackage{bm}
\usepackage{mathrsfs}
\usepackage{tabularx}
\usepackage{bigstrut}
\usepackage{epstopdf}
\usepackage{xfrac}
\usepackage{listings}
\usepackage{enumitem}
\usepackage{tensor}
\usepackage{dsfont}

\usepackage[usenames,dvipsnames]{xcolor}
\usepackage[hyperindex,pdftex, breaklinks,colorlinks = true,linkcolor = blue,urlcolor=blue,citecolor=blue]{hyperref}

\usepackage{caption}
\usepackage{subcaption}
\usepackage{physics}
\usepackage[section]{placeins}
\usepackage{tensor}
\usepackage{comment}

\usepackage{dsfont}

\begin{document}

\preprint{APS/123-QED}

\title{Multipolar quantum electrodynamics of localized charge-current distributions: \\
 Spectral theory and renormalization}

\author{Jason G. Kattan}
\email{jkattan@physics.utoronto.ca}
\affiliation{Department of Physics, University of Toronto, Toronto, Ontario M5S 1A7, Canada}

\author{J. E. Sipe}
\email{sipe@physics.utoronto.ca}
\affiliation{Department of Physics, University of Toronto, Toronto, Ontario M5S 1A7, Canada}

\date{\today}
\begin{abstract}
    We formulate a non-relativistic quantum field theory to model interactions between quantized electromagnetic fields and localized charge-current distributions. The electronic degrees of freedom are encoded in microscopic polarization and magnetization field operators whose moments are identified with the multipole moments of the charge-current distribution. The multipolar Hamiltonian is obtained from the minimal coupling Hamiltonian through a unitary transformation, often referred to as the Power-Zienau-Woolley transformation; we renormalize this Hamiltonian using perturbation theory, the result of which is used to compute the leading-order radiative corrections to the electronic energy levels due to interactions between the electrons and quantum vacuum fluctuations in the electromagnetic field. Our renormalized energy shift constitutes a generalization of the Lamb shift in atomic hydrogen, valid for general localized assemblies of atoms and molecules, possibly with net charge but absent free current. By expanding the fields in a series of multipole moments, our results can be used to study contributions to this energy shift coming from specific multipole moments of arbitrary order.
\end{abstract}
 
\maketitle

\section{Introduction}\label{introduction}

While relativistic quantum electrodynamics has been tremendously successful in describing scattering processes in high energy physics, it is cumbersome when applied to bound state problems and those with non-relativistic sources \cite{Caswell, Labelle}. In part, this is because any process involving matter has an antimatter counterpart that must be included in calculations of probability amplitudes, \textit{e.g.}, to preserve unitarity of the $\mathrm{S}$-matrix \cite{LandauLifshitz}. However, after taking the non-relativistic limit the antimatter degrees of freedom decouple in the path integral and can be integrated out of the theory \cite{Foldy, Burgess}. 

The result is an effective field theory, called \textit{non-relativistic quantum electrodynamics} \cite{Healy}, that is particularly well-suited for describing interactions between many-body systems with a \textit{fixed} number of particles and quantized radiation fields. It is obtained from an ultraviolet completion with only one characteristic energy scale (set by the electron mass), implying that the infinite sum of interaction terms in its Lagrangian can be ordered in reciprocal powers of this mass \cite{Paz1}; for applications in quantum optics and condensed matter physics, it is often sufficient to consider only the leading-order terms. In this Lagrangian, matter is described by charge and current densities $(\rho,\bm{j})$ that couple to the $\mathrm{U}(1)$--gauge potentials $(\phi,\bm{a})$ rather than directly to the electromagnetic field.

In many systems, however, an alternative approach is possible, provided one can identify specific points in space around which it is physically permissible to perform multipole expansions. This approach is based on a formalism called \textit{multipolar electrodynamics} \cite{AtkinsWoolley,Woolley6,babiker1,Woolley3}, which has been specifically designed to study such systems. In this paradigm, the material degrees of freedom are encoded in polarization and magnetization fields $(\bm{p},\bm{m})$ in place of charge and current densities $(\rho,\bm{j})$, which couple to the ``Maxwell fields" $(\bm{d},\bm{b})$ rather than the gauge potentials $(\phi,\bm{a})$. These descriptions are related by
\begin{align}
    \rho(\bm{x},t) &= - \bm{\nabla} \cdot \bm{p}(\bm{x},t) + \rho_{F}(\bm{x},t)\nonumber, \\
    \bm{j}(\bm{x},t) &= \frac{\partial \bm{p}(\bm{x},t)}{\partial t}  + c\bm{\nabla} \times \bm{m}(\bm{x},t) + \bm{j}_{F}(\bm{x},t),\nonumber
\end{align}
allowing for the possibility of free charge and current densities $(\rho_F, \bm{j}_F)$, which would arise if the atom or molecule had a net charge and was free to move \cite{MolQED}. In its Hamiltonian formulation, the standard procedure for obtaining the quantum theory of multipolar electrodynamics involves a unitary transformation of the minimal coupling Hamiltonian \cite{Woolley4,Dmytruk1}, often referred to as the Power-Zienau-Woolley (PZW) transformation after its originators \cite{PowerZienau, AtkinsWoolley, Woolley6}. Calculations based on the resulting \textit{multipolar Hamiltonian} often afford more physical insight into the dynamics of the system than do those based on the minimal coupling Hamiltonian, and are free of the artificial divergences in the determination of response coefficients that can plague the latter \cite{Sipe1}.

Multipolar electrodynamics was originally formulated in terms of the many-body wavefunctions of ``first-quantized" quantum mechanics \cite{Perspective}. While their use simplifies certain calculations, these wavefunctions become increasingly difficult to work with as the number of particles becomes large, and certainly prohibitively so if one envisions ultimately extending the treatment to condensed matter systems. Computations are vastly simplified when recast in terms of second-quantized field operators, which automatically account for the various combinatorial factors arising from particle exchange symmetry and permit the application of powerful Fock space methods from quantum field theory \cite{schweber}.

Even in second-quantized treatments, the multipolar Hamiltonian is usually simplified by making the ubiquitious ``electric dipole approximation" \cite{Salam1, Haugland1, Haugland2, Rokaj, Funai}, neglecting higher-order electric and magnetic multipole moments that become important when spatial variations of the electromagnetic field over the atom or molecule are non-negligible. And, indeed, for larger systems, such as molecules with several distinct chromophores, spatial variations in optical fields are non-negligible, and the electric dipole approximation is invalid \cite{MolQED}. Cases involving multipole moments beyond the electric dipole have been investigated previously \cite{SalamTh}, notably in calculations of intermolecular dispersion interactions \cite{SalamTh2} and resonance energy transfer rates \cite{Salam3}; however, the formulation of second-quantized models of radiation-matter interactions that are also capable of including electric and magnetic multipole moments \textit{of arbitrary order} has not yet been addressed.

In this paper we present such an approach. We reformulate multipolar electrodynamics within the framework of quantum field theory --- a formalism that we call \textit{multipolar quantum electrodynamics} --- using a field-theoretic generalization of the PZW transformation applied to the minimal coupling Hamiltonian. The electronic degrees of freedom are encoded in second-quantized scalar field operators that are used to define \textit{microscopic} polarization and magnetization field operators. These microscopic fields contain the full electric and magnetic multipole series, and couple directly to the quantized electromagnetic field through interaction terms of the same form as their first-quantized counterparts. While an analogous reformulation has been introduced previously \cite{babiker1, Salam2, PowerTh1}, the focus of those studies was the fields in the vicinity of an atom or molecule \cite{PowerTh2} and the impact of the associated local field corrections on intermolecular interactions \cite{PowerTh3, PowerTh4, PowerTh5}. Our focus instead concerns the vacuum structure of the theory, which is responsible for a vast landscape of phenomena in quantum electrodynamics including the Lamb shift \cite{Lamb}, spontaneous emission \cite{Milonni}, resonance energy transfer rates \cite{SalamTh3}, and the Casimir \cite{Casimir} and Casimir-Polder interactions \cite{SalamTh2, CasimirPolder}. 

In this first communication, we use perturbation theory to calculate the leading-order radiative corrections to the electronic energy levels resulting from interactions between the electrons in an isolated atom or molecule and quantum vacuum fluctuations of the electromagnetic field. Such perturbative calculations generically yield ultraviolet divergences, necessitating regularization and renormalization. We renormalize the multipolar Hamiltonian at leading-order in Rayleigh-Schrödinger perturbation theory, using standard techniques from effective field theory \cite{Burgess}. Our renormalized energy shift constitutes a generalization of Bethe's original calculation of the Lamb shift in atomic hydrogen \cite{Bethe}, valid for general localized assemblies of atoms and molecules, with or without net charge. In its multipolar form, this renormalized energy shift can be expanded in a sum of contributions coming from an arbitrary number of electric and magnetic multipole moments, so that vacuum effects can be studied order-by-order in the multipole series. 

We begin in Section \ref{section1} by recalling the essential features of minimal coupling electrodynamics. The minimal coupling Lagrangian is obtained directly from the Schrödinger Lagrangian by replacing partial derivatives with $\mathrm{U}(1)$--covariant derivatives and adding to the result a kinetic term yielding the free Maxwell equations \cite{Schwartz}. To obtain the Hamiltonian field theory, one employs Dirac's extended Hamiltonian formalism \cite{Diraclecs}, since the minimal coupling Lagrangian is degenerate and the associated Hamiltonian will therefore be subject to constraints. Briefly summarizing this standard analysis in Appendix \ref{appendixsum1}, we obtain the quantized minimal coupling Hamiltonian, together with the commutation/anticommutation relations for the field operators. Applying the unitary PZW transformation to the canonical variables and Hamiltonian of minimal coupling (summarized in Appendix \ref{appendixsum2}), the quantized Hamiltonian theory of multipolar electrodynamics follows.

In Section \ref{section2} we present a sketch of the procedure we follow to regularize and renormalize the multipolar Hamiltonian, and in Section \ref{section3} we compute the regularized energy shift of the electronic energy levels in the electromagnetic vacuum state. We regularize the divergent integrals with a hard cutoff $\|\bm{k}\|\leq \Lambda$, introducing a second energy scale into the theory. However, this scale is artificial and should not appear in expressions for measurable quantities; to remove this $\Lambda$-dependence from our energy shift, we renormalize the multipolar Hamiltonian following the procedure outlined in Section \ref{section2}. This is done in Section \ref{section4}, the result of which is a finite and observable shift of the electronic energy levels. We confirm that for a single electron atom this energy shift reduces to Bethe's result for the Lamb shift in the electric dipole approximation \cite{Bethe}, and its generalizations when spatial variations in the electromagnetic field over the atom are included agrees with the results of other authors \cite{Au, Grotch} who extended Bethe's result in this direction. We conclude by expanding this renormalized energy shift in a sum of contributions coming from the first few electric and magnetic multipole moments, and quote the expressions for higher-order multipole contributions. We summarize our results in Section \ref{section5}, and some of the details are relegated to the Appendices.

\section{Hamiltonian Theory}\label{section1}
\subsection{Minimal coupling}\label{subsection1A}
We consider a system of electrons interacting with one or more point-like ions at fixed positions clustered around a specific point $\bm{R} \in \mathbb{R}^3$. Denoting by $q_{N}$ the charge of the $N^{\text{th}}$ ion located at $\bm{d}_N$ with respect to $\bm{R}$, the charge density of the ions is 
\begin{equation}
    \rho^{\text{ion}}(\bm{x}) = \sum\limits_{N} q_{N}\, \delta(\bm{x} - \bm{R} - \bm{d}_N).
    \label{eq2a1}
\end{equation}
The ionic charge density $\rho^{\text{ion}}(\bm{x})$ leads to an electrostatic interaction between the ions that we ignore under the frozen-ion approximation, together with a fixed background potential through which the electrons propagate of the form 
\begin{equation}
    \mathrm{V}(\bm{x}) = e \sum\limits_{N} \frac{q_{N}}{\|\bm{x} - \bm{R} - \bm{d}_N\|},
    \label{eq2a2}
\end{equation}
where $\|\bm{x}\| = \sqrt{\bm{x}\cdot\bm{x}}$ is the Euclidean $2$-norm and $e = - \abs{e}$ is the charge of an electron. The dynamics of non-interacting electrons propagating within this background potential are described as usual by the Schrödinger equation, which in the coordinate representation can be derived from the Lagrangian density
\begin{align}
    \mathcal{L}_F &= \frac{i \hbar}{2} \Big(\mathcal{\psi}^{\dagger}(\bm{x},t) \dot{\psi}(\bm{x},t) - \dot{\psi}^{\dagger}(\bm{x},t)\psi(\bm{x},t)\Big)\nonumber \\
    &- \frac{\hbar^2}{2m} \bm{\nabla}\psi^{\dagger}(\bm{x},t)\cdot\bm{\nabla}\psi(\bm{x},t) - \psi^{\dagger}(\bm{x},t) \mathrm{V}(\bm{x}) \psi(\bm{x},t).
    \label{eq2a3}
\end{align}

To couple the electron field $\psi(\bm{x},t)$ to a classical electromagnetic field, we apply the standard minimal coupling prescription to the Lagrangian density above \cite{Schwartz}. In the four-vector notation of relativistic mechanics, this amounts to the replacement
\begin{equation}
    \partial_{\mu} \to \partial_{\mu} + \frac{ie}{\hbar c} a_{\mu}(\bm{x},t) \equiv \mathcal{D}_{\mu}(\bm{x},t),
    \label{cov}
\end{equation}
where the object on the right is a $\mathrm{U}(1)$--covariant derivative and $a_{\mu}(\bm{x},t)$ is the electromagnetic four-potential. And to give dynamics to the electromagnetic field we add to the result a kinetic term
\begin{equation}
    \mathcal{L}_B = \frac{1}{8\pi} \Big(\bm{e}(\bm{x},t)\cdot\bm{e}(\bm{x},t) - \bm{b}(\bm{x},t)\cdot\bm{b}(\bm{x},t)\Big)
    \label{eq2temp1}
\end{equation}
associated with the \textit{free} Maxwell equations. The $\mathrm{U}(1)$--gauge potentials $(\phi(\bm{x},t),\bm{a}(\bm{x},t))$ are related to the electric and magnetic fields $(\bm{e}(\bm{x},t),\bm{b}(\bm{x},t))$ through the usual relations \cite{Jackson}
\begin{align}
    \bm{e}(\bm{x},t) &= - \bm{\nabla} \phi(\bm{x},t) - \frac{1}{c} \frac{\partial \bm{a}(\bm{x},t)}{\partial t},\label{eq2a13} \\
    \bm{b}(\bm{x},t) &= \bm{\nabla} \times \bm{a}(\bm{x},t).\label{eq2a14}
\end{align}
Then the total Lagrangian density for non-relativistic quantum electrodynamics in minimal coupling is 
\begin{widetext}
\begin{align}
    \mathcal{L} = \frac{1}{8\pi} \Big(\bm{e}(\bm{x},t)\cdot\bm{e}(\bm{x},t) - \bm{b}(\bm{x},t)\cdot\bm{b}(\bm{x},t)\Big) + \frac{i\hbar}{2}\Big(\psi^{\dagger}(\bm{x},t)\dot{\psi}(\bm{x},t) - \dot{\psi}^{\dagger}(\bm{x},t)\psi(\bm{x},t)\Big) - \psi^{\dagger}(\bm{x},t)\mathrm{V}(\bm{x})\psi(\bm{x},t)\nonumber \\
    - \frac{\hbar^2}{2m} \big(\bm{\mathcal{D}}(\bm{x},t)\psi(\bm{x},t)\big)^{\dagger}\cdot\big(\bm{\mathcal{D}}(\bm{x},t)\psi(\bm{x},t)\big)
     - \rho^e(\bm{x},t) \phi(\bm{x},t),
\end{align}
\end{widetext}
where $\mathcal{D}^i(\bm{x},t)$ are the spatial components of the covariant derivative defined in (\ref{cov}) and we have introduced the electronic charge density
\begin{equation}
    \rho^e(\bm{x},t) = e\psi^{\dagger}(\bm{x},t) \psi(\bm{x},t).
\end{equation}
Expanding the covariant derivative, we can write the total Lagrangian associated to $\mathcal{L}$ as a sum of three terms
\begin{equation}
    L = L_{B} + L_{F} + L_{\text{int}},\label{eq2T1}
\end{equation}
where $L_B$ and $L_F$ are integrals over $\mathbb{R}^3$ of the Lagrangian densities given in (\ref{eq2temp1}) and (\ref{eq2a3}), and the interaction term is
\begin{widetext}
\begin{align}
    L_{\text{int}} = - \int d\bm{x}\,\rho^e(\bm{x},t)\phi(\bm{x},t) + \frac{\hbar e}{2mc i}\int d\bm{x} \Big(\psi^{\dagger}(\bm{x},t)\bm{\nabla}\psi(\bm{x},t) - \bm{\nabla}\psi^{\dagger}(\bm{x},t)\psi(\bm{x},t)\Big)\cdot\bm{a}(\bm{x},t)\nonumber \\
    - \frac{e^2}{2mc^2} \int d\bm{x}\,\psi^{\dagger}(\bm{x},t) \|\bm{a}(\bm{x},t)\|^2 \psi(\bm{x},t).
\end{align}
\end{widetext}

To obtain the Hamiltonian field theory we apply the Legendre transformation to $L$, which is summarized in Appendix \ref{appendixsum1}. The Hamiltonian operator that we obtain from $L$ after quantization is given in the Schrödinger picture by
\begin{equation}
    H = H_B \otimes \mathbb{I}_F + \mathbb{I}_B \otimes H_F + H_{\text{int}}
\label{eq2a19}
\end{equation}
generating unitary dynamics in a composite Hilbert space $\mathcal{H}_{B} \otimes \mathcal{H}_F$ constructed from Hilbert spaces $\mathcal{H}_B$ and $\mathcal{H}_F$ associated with the Bose and Fermi sectors, respectively. In terms of the \textit{transverse} electric and magnetic field operators, the Maxwell Hamiltonian is
\begin{equation}
    H_B = \frac{1}{8\pi} \int d\bm{x}\, \Big(\bm{e}_T(\bm{x}) \cdot \bm{e}_T(\bm{x}) + \bm{b}(\bm{x})\cdot\bm{b}(\bm{x})\Big),\label{temp11}
\end{equation}
\\
the electronic term is
\begin{align}
    H_F = \frac{\hbar^2}{2m} \int d\bm{x}\,\bm{\nabla}\psi^{\dagger}(\bm{x})\cdot\bm{\nabla}\psi(\bm{x}) + \frac{1}{2}\int d\bm{x}\, \rho^e(\bm{x})\phi(\bm{x})\nonumber \\
    + \int d\bm{x}\, \psi^{\dagger}(\bm{x})\mathrm{V}(\bm{x})\psi(\bm{x}),
\end{align}
and the interaction term is
\begin{equation}
    H_{\text{int}} = -\frac{1}{c} \int d\bm{x}\, \bm{j}_P(\bm{x})\cdot\bm{a}(\bm{x}) - \frac{1}{2c} \int d\bm{x}\, \bm{j}_D(\bm{x})\cdot\bm{a}(\bm{x}).\label{eq28}
\end{equation}
Here we have introduced the paramagnetic and diamagnetic current density operators
\begin{align}
    \bm{j}_P(\bm{x}) &= \frac{\hbar e}{2mi}\Big(\psi^{\dagger}(\bm{x})\bm{\nabla}\psi(\bm{x}) - \bm{\nabla}\psi^{\dagger}(\bm{x})\psi(\bm{x})\Big),\label{minpara}\\
    \bm{j}_D(\bm{x}) &= - \frac{e^2}{mc} \psi^{\dagger}(\bm{x})\bm{a}(\bm{x})\psi(\bm{x}),\label{mindia}
\end{align}
such that the total current density operator is
\begin{equation}
    \bm{j}(\bm{x}) = \bm{j}_P(\bm{x}) + \bm{j}_D(\bm{x}),
\end{equation}
which satisfies the Heisenberg-picture equation
\begin{equation}
    \frac{\partial \rho^e(\bm{x},t)}{\partial t} + \bm{\nabla}\cdot\bm{j}(\bm{x},t) = 0
\end{equation}
associated with local $\mathrm{U}(1)$--charge conservation. The electromagnetic field operators satisfy the (equal-time) commutation relations
\begin{align}
    \big[e_T^i(\bm{x}), a^j(\bm{y})\big]_{-} &= 4\pi i \hbar c\, \delta^{ij}_{T}(\bm{x}-\bm{y}),\nonumber \\
    \big[a^i(\bm{x}), a^j(\bm{y})\big]_{-} &= \big[e_T^i(\bm{x}), e_T^j(\bm{y})\big]_{-} = 0,\label{eq2a17}
\end{align}
from which follows
\begin{equation}
    \big[e_T^i(\bm{x}), b^j(\bm{y})\big]_{-} = 4\pi i \hbar c\, \varepsilon^{ijk}\frac{\partial}{\partial y^k} \delta(\bm{x} - \bm{y})\, ,\label{eq2a18}
\end{equation} 
while the electron field operator and its adjoint satisfy the (equal-time) anticommutation relations
\begin{align}
    \big[\psi(\bm{x}), \psi^{\dagger}(\bm{y})\big]_+ &= \delta(\bm{x} - \bm{y}),\nonumber \\
    \big[\psi(\bm{x}), \psi(\bm{y})\big]_+ &= \big[\psi^{\dagger}(\bm{x}), \psi^{\dagger}(\bm{y})\big]_+ = 0.
    \label{eq2a10}
\end{align}
In the commutation relations (\ref{eq2a17}) we have introduced the transverse delta function $\delta^{ij}_T(\bm{x}-\bm{y})$, which is used together with the longitudinal delta function $\delta^{ij}_{L}(\bm{x}-\bm{y})$ to decompose a given vector field $\mathcal{O}^i(\bm{x})$ into its transverse and longitudinal components \cite{MolQED}
\begin{align}
    \mathcal{O}_T^i(\bm{x}) &= \int d\bm{y}\, \delta^{ij}_{T}(\bm{x}-\bm{y}) \mathcal{O}^j(\bm{y}), \\
    \mathcal{O}_{L}^i(\bm{x}) &=  \int d\bm{y}\, \delta^{ij}_{L}(\bm{x}-\bm{y}) \mathcal{O}^j(\bm{y}).
\end{align}

\subsection{Multipolar quantum electrodynamics}\label{subsection1B}
In classical or semiclassical theory, the multipolar Hamiltonian is obtained 
from the minimal coupling Hamiltonian through a canonical transformation. After the transformation, the Hamiltonian involves the electric and magnetic fields rather than the scalar and vector potentials, together with \textit{polarization} and \textit{magnetization} fields that are the microscopic analogues of the polarization and magnetization fields appearing in elementary classical electrodynamics \cite{Jackson}. The moments of the polarization and magnetization fields are identified with the multipole moments of the charge-current distribution.

In the fully quantum theory, instead of a canonical transformation there is a corresponding unitary transformation, called the PZW transformation \cite{Perspective}, that acts on the composite Hilbert space $\mathcal{H}_B \otimes \mathcal{H}_F$, generated by
\begin{equation}
    S = \frac{1}{c} \int d\bm{x}\, \bm{p}(\bm{x}) \cdot \bm{a}(\bm{x}),\label{eq2b1}
\end{equation}
 where $\bm{p}(\bm{x})$ is the polarization field operator, which is related to the total charge density operator
\begin{equation}
    \rho(\bm{x}) = \rho^e(\bm{x}) + \rho^{\text{ion}}(\bm{x})
\end{equation}
through the identity
\begin{equation}
    \rho(\bm{x}) = - \bm{\nabla} \cdot\bm{p}(\bm{x}) + \rho_F(\bm{x}).\label{eq2b3} 
\end{equation}
The free charge density $\rho_F(\bm{x})$ is nonzero only if the system has net charge, in which case
\begin{equation}
    \rho_F(\bm{x}) = Q\, \delta(\bm{x}-\bm{R})
    \label{eq2b4}
\end{equation}
with net charge 
\begin{equation}
    Q = \int d\bm{x}\, \rho^e(\bm{x}) + \sum\limits_{N} q_{N}.
    \label{eq2b5}
\end{equation}
We allow $Q$ to be non-zero to include single ions or charged molecules. A suitable polarization field operator that satisfies (\ref{eq2b3}) is
\begin{eqnarray}
    \bm{p}(\bm{x}) = \int d\bm{y}\, \bm{s}(\bm{x};\bm{y},\bm{R}) \rho(\bm{y}),
    \label{eq2b6}
\end{eqnarray}
where we have introduced a ``relator" $\bm{s}(\bm{x};\bm{y},\bm{R})$ \cite{Healy}, defined by the distributional expression 
\begin{equation}
    \bm{s}(\bm{x};\bm{y},\bm{R}) = \int_{C(\bm{y},\bm{R})} d\bm{z}\, \delta(\bm{x} - \bm{z})
    \label{eq2b7}
\end{equation}
with $C(\bm{y},\bm{R})$ an arbitrary continuously differentiable curve in $\mathbb{R}^3$ that begins at $\bm{R}$ and ends at $\bm{y}$. If a straight-line path is chosen between $\bm{R}$ and $\bm{y}$, then the usual multipole expansion of the polarization field follows.

After applying the PZW transformation, which is summarized in Appendix \ref{appendixsum2}, the multipolar Hamiltonian that follows is
\begin{equation}
    H = H_B \otimes \mathbb{I}_F + \mathbb{I}_B \otimes H_F + H_{\text{int}},
    \label{temp1}
\end{equation}
where $H_B$ is now written in terms of the transverse displacement field instead of the transverse electric field
\begin{equation}
    H_B = \frac{1}{8\pi} \int d\bm{x}\, \Big(\bm{d}_T(\bm{x})\cdot\bm{d}_T(\bm{x}) + \bm{b}(\bm{x})\cdot \bm{b}(\bm{x})\Big),
    \label{temp2}
\end{equation}
the electronic term is
\begin{align}
    H_F = \frac{\hbar^2}{2m} \int d\bm{x}\, \bm{\nabla}\psi^{\dagger}(\bm{x})\cdot\bm{\nabla}\psi(\bm{x}) + 2\pi \int d\bm{x}\, \|\bm{p}_{L}(\bm{x})\|^2\nonumber \\
    - \int d\bm{x}\, \bm{d}_{L}(\bm{x})\cdot\left(\bm{p}(\bm{x}) - \frac{1}{8\pi} \bm{d}_{L}(\bm{x})\right),
    \label{temp3}
\end{align}
and the interaction term is
\begin{align}
    H_{\text{int}} &= 2\pi \int d\bm{x}\, \|\bm{p}_T(\bm{x})\|^2  - \int d\bm{x}\, \bm{p}(\bm{x})\cdot\bm{d}_T(\bm{x})\nonumber \\
    &- \int d\bm{x}\, \bm{m}_P(\bm{x})\cdot\bm{b}(\bm{x}) - \frac{1}{2} \int d\bm{x}\, \bm{m}_D(\bm{x})\cdot\bm{b}(\bm{x}).
    \label{temp4}
\end{align}
Importantly, here and henceforth the electron field operator $\psi(\bm{x})$ is the \textit{transformed} operator (see Appendix \ref{appendixsum2}). The longitudinal displacement field, defined through Gauss's law, $\bm{\nabla} \cdot\bm{d}_{L}(\bm{x}) = 4\pi\rho_{F}(\bm{x})$, is given by
\begin{equation}
    \bm{d}_{L}(\bm{x}) = \frac{1}{4\pi} \int d\bm{y}\, \rho_F(\bm{y}) \left(\frac{\hat{\bm{x}}-\hat{\bm{y}}}{\|\bm{x}-\bm{y}\|^2}\right).
\end{equation}

In the interaction term (\ref{temp4}) $\bm{m}_P(\bm{x})$ and $\bm{m}_D(\bm{x})$ refer to the paramagnetic and diamagnetic contributions to the (orbital) magnetization field operator
\begin{equation}
    \bm{m}(\bm{x}) = \bm{m}_P(\bm{x}) + \bm{m}_D(\bm{x}).
\end{equation}
Explicitly, the field operators $\bm{m}_{P}(\bm{x})$ and $\bm{m}_D(\bm{x})$ are
\begin{align}
    m_{P}^i(\bm{x}) &= \frac{1}{c} \int d\bm{y}\, \alpha^{ij}(\bm{x}; \bm{y},\bm{R}) j_{P}^{j}(\bm{y}),\label{eq2b27} \\
    m_{D}^i(\bm{x}) &= \frac{1}{c} \int d\bm{y}\, \alpha^{ij}(\bm{x}; \bm{y},\bm{R}) j_{D}^{j}(\bm{y}),\label{eq2b28}
\end{align}
where the paramagnetic and diamagnetic current density operators are 
\begin{align}
    \bm{j}_{P}(\bm{x}) &= \frac{\hbar e}{2mi} \Big(\psi^{\dagger}(\bm{x}) \bm{\nabla} \psi(\bm{x}) - \bm{\nabla}\psi^{\dagger}(\bm{x}) \psi(\bm{x})\Big),\label{eq2b29} \\
    \bm{j}_{D}(\bm{x}) &= - \frac{e^2}{m c} \psi^{\dagger}(\bm{x}) \bm{\Omega}_{\bm{R}}(\bm{x}) \psi(\bm{x}),\label{eq2b30}
\end{align}
and we have introduced another ``relator" \cite{Healy}
\begin{equation}
    \alpha^{ij}(\bm{x};\bm{y},\bm{R}) = \varepsilon^{i m n} \int_{C(\bm{y},\bm{R})} dz^m\, \frac{\partial z^n}{\partial y^{j}} \delta(\bm{x} - \bm{z}).
    \label{eq2b31}
\end{equation}
As is the case for the polarization field (\ref{eq2b6}), taking a straight-line path for $C(\bm{y},\bm{R})$ yields the multipole expansion of the magnetization fields (\ref{eq2b27},\ref{eq2b28}). In multipolar electrodynamics, the diamagnetic current density (\ref{eq2b30}) depends not on the vector potential, but instead on the magnetic field through
\begin{equation}
    \Omega_{\bm{R}}^i(\bm{x}) = \int d\bm{y}\, \alpha^{ji}(\bm{y}; \bm{x},\bm{R}) b^{j}(\bm{y}).
    \label{eq2b32}
\end{equation}
The charge and current density operators are related to the polarization and magnetization field operators through the Heisenberg-picture identities
\begin{align}
    \rho(\bm{x},t) &= - \bm{\nabla} \cdot \bm{p}(\bm{x},t) + \rho_F(\bm{x}),\label{temp5} \\
    \bm{j}(\bm{x},t) &= \frac{\partial \bm{p}(\bm{x},t)}{\partial t} + c\bm{\nabla} \times \bm{m}(\bm{x},t)\label{temp6}.
\end{align}

\section{Renormalization Strategy}\label{section2}
The composite Hilbert space of multipolar quantum electrodynamics is the tensor product $\mathcal{H}_{B} \otimes \mathcal{H}_{F}$ of the Hilbert spaces $\mathcal{H}_B$ and $\mathcal{H}_F$ associated with the Bose and Fermi sectors, respectively. Since implicit in the Hamiltonian $H$ is the assumption that the electron particle number is conserved, we introduce an ultraviolet (UV) cutoff $\Lambda$, excluding contributions to Fourier integrals coming from electromagnetic field modes of $\|\bm{k}\| > \Lambda$. We
denote by $H^\Lambda$ our Hamiltonian $H$ subject to this cutoff, which we will henceforth refer to as the \textit{regularized Hamiltonian}, and using the same superscript to denote regularized versions of the different contributions to $H$. Choosing $\Lambda$ to be on the order of the inverse (reduced) Compton wavelength of the electron, this guarantees that we are restricting ourselves to energies for which the electron-positron pair production that we neglect would not be present in the full theory of QED \cite{Feynman1}.

In the next section we will use Rayleigh-Schrödinger perturbation theory to compute the leading-order radiative corrections to the electronic energy levels resulting from interactions between the electrons and quantum vacuum fluctuations in the electromagnetic field. Were $\Lambda$ taken to infinity, the typical ultraviolet divergences would result. With the ``hard cutoff" imposed each regularized energy shift, denoted by $\Delta E_n^{\Lambda}$, is finite, albeit explicitly dependent on $\Lambda$. However, this dependence on $\Lambda$ is unphysical and should not appear in observable quantities; it arises because $\Delta E_n^{\Lambda}$ contains an unobservable, $\Lambda$-dependent contribution coming from \textit{free electrons} interacting with the electromagnetic vacuum. Since this contribution is unobservable, we should subtract it from $\Delta E_n^{\Lambda}$; what remains will be the observable correction to the \textit{bound state} electronic energy levels.


This can be implemented by adding one or more local interactions $\delta H(\Lambda)$, depending explicitly on $\Lambda$, to the regularized Hamiltonian $H^{\Lambda}$, with the purpose of simulating virtual processes associated with the high energy ($\|\bm{k}\| > \Lambda$) degrees of freedom \cite{Lepage}. In general, $\delta H(\Lambda)$ will be a sum of many terms, each summand being called a \textit{counterterm}. A central tenet of renormalization theory is that it should be possible to 
choose the counterterms in $\delta H(\Lambda)$ such that the \textit{renormalized Hamiltonian}
\begin{equation}
    H^R = H^{\Lambda} + \delta H(\Lambda)
\end{equation}
accurately represents the full range of energies in the interval $0 \leq \|\bm{k}\| < \infty$ and should \textit{not} depend on $\Lambda$ \cite{GlazekWilson}. The \textit{renormalized} energy shift, denoted by $\Delta E_n^R$ and computed using $H^R$, should be finite in the limit $\Lambda \to \infty$ and represents the observable correction to the electronic energy levels in the electromagnetic vacuum. 

This is the usual strategy we implement here. However, values of $\|\bm{k}\|$ above the inverse (reduced) Compton wavelength of the electron are treated improperly by the theory, and so extensions of $\Lambda$ greater than that go beyond the validity of the Hamiltonian. And, indeed, non-relativistic QED is a \textit{non-renormalizable} field theory, so we only expect $\Delta E_n^R$ to be finite to the order of perturbation theory at which we renormalized the Hamiltonian, since additional divergences are generated in non-renormalizable theories as one moves to higher orders, as well as divergences that are generated by the counterterms in $\delta H(\Lambda)$ \cite{Schwartz}. If we want to compute observables at next-to-leading-order, for example, we must check if additional UV divergences are generated. In this paper we will only renormalize the multipolar Hamiltonian at leading-order in the fine structure constant.

\section{Energy Shifts}\label{section3}

\subsection{Preliminaries}\label{subsection3A}
We begin by identifying the contributions to the energy shifts arising from the free fields. For the radiation field this is standard, but we take the opportunity to introduce notation that we will use later in the paper; for the electrons the computation is done in terms of fermionic scalar fields rather than the single or few-electron states that are typically used. Finally, we identify the two contributions that arise from the interaction of the radiation and electron fields. They are then worked out in detail in Sections \ref{subsection3C} and \ref{subsection3D} below.

\subsubsection{Free fields}
The dynamical evolution of the transverse Maxwell fields $\bm{d}_T(\bm{x},t)$ and $\bm{b}(\bm{x},t)$ is generated by $H_{B}$ through the Heisenberg equations
\begin{align}
    \frac{\partial\bm{d}_T(\bm{x},t)}{\partial t} &= \frac{1}{i\hbar} \big[\bm{d}_T(\bm{x},t), H_B\big]_-,\label{eq3a1} \\
    \frac{\partial\bm{b}(\bm{x},t)}{\partial t} &= \frac{1}{i\hbar} \big[\bm{b}(\bm{x},t), H_B\big]_-,\label{eq3a2}
\end{align}
which, with the explicit form (\ref{temp2}) for $H_B$, lead to the familiar Maxwell equations
\begin{align}
    \frac{\partial\bm{d}_T(\bm{x},t)}{\partial t} &= + c\bm{\nabla} \times \bm{b}(\bm{x},t),\label{eq3a3} \\
    \frac{\partial\bm{b}(\bm{x},t)}{\partial t} &= - c\bm{\nabla} \times \bm{d}_T(\bm{x},t),\label{eq3a4}
\end{align}
subject to the constraints
\begin{align}
    \bm{\nabla} \cdot \bm{d}_T(\bm{x},t) &= 0,\label{eq3a5} \\
    \bm{\nabla} \cdot \bm{b}(\bm{x},t) &= 0.\label{eq3a6}
\end{align}
We quantize in a box of volume $V = L^3$, in which case the allowed wavevectors are
\begin{equation}
    \bm{k} = \frac{2\pi}{L} \big(n_x, n_y, n_z\big) \in \frac{2\pi}{L} \mathbb{Z}^3,
\end{equation}
and expand $\bm{d}_T(\bm{x},t)$ and $\bm{b}(\bm{x},t)$ in a Fourier series of transverse spatial modes,  which can themselves be written as a sum of $\mathbb{C}$-valued circular polarization vectors $\bm{e}_{I\bm{k}}$ ($I = L,R$) defined by
\begin{align}
    \bm{e}_{L\bm{k}} &= -\frac{1}{\sqrt{2}} \left(\bm{\varepsilon}_{1\bm{k}} + i \bm{\varepsilon}_{2\bm{k}}\right),\nonumber \\
    \bm{e}_{R\bm{k}} &= +\frac{1}{\sqrt{2}} \left(\bm{\varepsilon}_{1\bm{k}} - i \bm{\varepsilon}_{2\bm{k}}\right),
\end{align}
where $\bm{\varepsilon}_{1\bm{k}}$ and $\bm{\varepsilon}_{2\bm{k}}$ are the standard Cartesian polarization vectors \cite{MolQED}. The circular polarization vectors satisfy
\begin{align}
    \bm{e}_{I\bm{k}}\cdot\bm{e}_{J(-\bm{k})} &= \delta_{IJ},\label{eq3a9}\\
    i\hat{\bm{k}}\times\bm{e}_{I\bm{k}} &= s_I \bm{e}_{I\bm{k}},\label{eq3a10}
\end{align}
where $s_L = + 1$ and $s_R = - 1$, and we choose the convention $\bm{e}_{I(-\bm{k})}= \bm{e}_{I\bm{k}}^*$. Moreover, these vectors satisfy the polarization sum rule
\begin{equation}
    \sum_{I} e_{I(-\bm{k})}^i e_{I\bm{k}}^j = \delta^{ij} - \frac{k^i k^j}{\|\bm{k}\|^2} \equiv \delta_T^{ij}(\bm{k}).\label{tdelta}
\end{equation}
The Maxwell field operators are
\begin{align}
    \bm{d}_T(\bm{x},t) &= i \sum\limits_{I\bm{k}} \left(\frac{2\pi\hbar \omega_{\bm{k}}}{V}\right)^{1/2} \bm{e}_{I\bm{k}} a_{I\bm{k}} e^{i(\bm{k}\cdot\bm{x} - \omega_{\bm{k}} t)} + \text{h.c.},\nonumber \\
    \bm{b}(\bm{x},t) &=  \sum\limits_{I\bm{k}} \left(\frac{2\pi\hbar \omega_{\bm{k}}}{V}\right)^{1/2} s_I \bm{e}_{I\bm{k}} a_{I\bm{k}} e^{i(\bm{k}\cdot\bm{x} - \omega_{\bm{k}} t)} + \text{h.c.},\label{eq3a13}
\end{align}
where $\omega_{\bm{k}} = c\|\bm{k}\|$. To enforce the commutation relations (\ref{eq2a17},\ref{eq2a18}), the photonic creation and annihilation operators $a_{I\bm{k}}^{\dagger}$ and $a_{I\bm{k}}$ must obey the equal-time canonical commutation relations
\begin{align}
    \big[a_{I\bm{k}}, a_{J\bm{k}'}^{\dagger}\big]_- &= \delta_{IJ} \delta_{\bm{k}\bm{k}'},\label{eq3a17} \\
    \big[a_{I\bm{k}}, a_{J\bm{k}'}\big]_- &= \big[a_{I\bm{k}}^{\dagger}, a_{J\bm{k}'}^{\dagger}\big]_- = 0.\label{eq3a18}
\end{align}

After regularization, the Maxwell Hamiltonian $H_B$ is given in the Schrödinger picture by
\begin{equation}
    H_{B}^{\Lambda} = \sum\limits_{I}\sum\limits_{\|\bm{k}\|\leq \Lambda} \hbar \omega_{\bm{k}} \left(a_{I\bm{k}}^{\dagger} a_{I\bm{k}} + \frac{1}{2}\right).
\end{equation}
The second term is the well-known divergence associated with the zero-point energy in free quantum electrodynamics \cite{Milonni, Schwartz}. Summing over polarization states, the energy of the vacuum state $\ket{\text{vac}} \in \mathcal{H}_B$ is exactly equal to this zero-point energy
\begin{equation}
    E_{0,B}^{\Lambda} \equiv \sum_{\|\bm{k}\|\leq \Lambda} \hbar c \|\bm{k}\| \;\to\; \frac{\hbar c V}{8\pi^2} \Lambda^4, \label{eq3T1}
\end{equation}
where the right side follows in the \textit{continuum limit}
\begin{equation}
    \sum_{\bm{k}} \to \frac{V}{8\pi^3} \int d\bm{k}.
    \label{cont}
\end{equation}
Therefore, the regularized Maxwell Hamiltonian is
\begin{equation}
    H_{B}^{\Lambda} = \sum\limits_{I}\sum\limits_{\|\bm{k}\|\leq \Lambda} \hbar \omega_{\bm{k}} a^{\dagger}_{I\bm{k}}a_{I\bm{k}} + E_{0,B}^{\Lambda}.\label{temp15}
\end{equation}
Since we are primarily interested in vacuum effects, we take as our basis for $\mathcal{H}_B$ the photon number states $\ket{\{n_{I\bm{k}}\}} \in \mathcal{H}_B$. The spectral problem for $H_{B}^{\Lambda}$ in terms of these eigenstates is
\begin{equation}
    H_{B}^{\Lambda} \ket{\{n_{I\bm{k}}\}} = E_B^{\Lambda} \ket{\{n_{I\bm{k}}\}},
\end{equation}
where the energy eigenvalues are
\begin{equation}
    E_{B}^{\Lambda} = \sum\limits_{I} \sum\limits_{\|\bm{k}\|\leq \Lambda} \hbar \omega_{\bm{k}} n_{I\bm{k}} + E_{0,B}^{\Lambda}.\label{temp17}
\end{equation}

Turning to the electron field, time evolution in the electronic Hilbert space $\mathcal{H}_F$ is generated by $H_{F}$ through the Heisenberg equation
\begin{equation}
    \frac{\partial\psi(\bm{x},t)}{\partial t} = \frac{1}{i\hbar} \big[\psi(\bm{x},t), H_F\big]_-,\label{eq3a19}
\end{equation}
with $H_F$ given by (\ref{temp3}). As shown in Appendix \ref{appendixA}, the regularized electronic Hamiltonian $H_F^{\Lambda}$ is given in the Schrödinger picture by
\begin{align}
    H_F^{\Lambda} &= \frac{\hbar^2}{2m}\int d\bm{x}\, \bm{\nabla}\psi^{\dagger}(\bm{x})\cdot\bm{\nabla}\psi(\bm{x})\nonumber \\
    &+ \frac{1}{2}\iint d\bm{x}d\bm{x}'\, \psi^{\dagger}(\bm{x})\psi^{\dagger}(\bm{x}')\frac{e^2}{\|\bm{x}-\bm{x}'\|}\psi(\bm{x}')\psi(\bm{x})\nonumber \\
    &+ \int d\bm{x}\, \psi^{\dagger}(\bm{x})\mathrm{V}(\bm{x})\psi(\bm{x}) + E_{0,F}^{\Lambda},\label{temp14}
\end{align}
where $E_{0,F}^{\Lambda}$ is a $\Lambda$-dependent term associated with the $\|\bm{x}-\bm{x}'\|\to 0$ limit of the electrostatic Coulomb potential, and is given by 
\begin{equation}
        E_{0,F}^{\Lambda} = \frac{1}{\pi} \left[e^2 N_e + 2 e N_e \sum_{N} q_N +\sum_N q_N^2\right]\Lambda,
\end{equation}
where $N_e$ is the total number of electrons. We take as our basis for $\mathcal{H}_F$ the set $\{\ket{\psi_n}\}_n$ of many-body eigenstates  of $H_F^{\Lambda}$. The spectral problem for $H_F^{\Lambda}$ in terms of these eigenstates is
\begin{equation}
    H_F^{\Lambda} \ket{\psi_n} = E_n^{\Lambda}\ket{\psi_n},\label{eq3b5}
\end{equation}
where the energy eigenvalues are
\begin{align}
    E_n^{\Lambda} &= \frac{\hbar^2}{2m}\int d\bm{x} \expval{\bm{\nabla}\psi^{\dagger}(\bm{x})\cdot\bm{\nabla}\psi(\bm{x})}_n\nonumber \\
    &+ \frac{1}{2}\iint d\bm{x}d\bm{x}' \expval{\psi^{\dagger}(\bm{x})\psi^{\dagger}(\bm{x}')\frac{e^2}{\|\bm{x}-\bm{x}'\|}\psi(\bm{x}')\psi(\bm{x})}_n\nonumber \\
    &+ \int d\bm{x} \expval{\psi^{\dagger}(\bm{x})\mathrm{V}(\bm{x})\psi(\bm{x})}_n + E_{0,F}^{\Lambda},\label{temp16}
\end{align}
and where the expectation value is over the state $\ket{\psi_n}$. 

\subsubsection{Interactions}
In the interacting theory the dynamical evolution of the field operators is generated by the full Hamiltonian $H$ given in (\ref{temp1}-\ref{temp4}). After regularization, we write this Hamiltonian as
\begin{equation}
    H^{\Lambda} = H_0^{\Lambda} + H_{\text{int}}^{\Lambda},\label{eq3a26}
\end{equation}
where
\begin{equation}
    H_0^{\Lambda} = H_B^{\Lambda} \otimes \mathbb{I}_F + \mathbb{I}_B \otimes H_F^{\Lambda} \label{eq3a27}
\end{equation}
with $H_B^{\Lambda}$ and $H_F^{\Lambda}$ given in (\ref{temp15}) and (\ref{temp14}), while $H_{\text{int}}^{\Lambda}$ is given by (\ref{temp4}) with all of the Fourier series (implicit in the mode expansions (\ref{eq3a13})) subject to the cutoff $\|\bm{k}\|\leq \Lambda$. 

We take as our basis for the composite Hilbert space $\mathcal{H}_B\otimes\mathcal{H}_F$ the product states 
\begin{equation}
    \ket{\{n_{I\bm{k}}\}; \psi_n} \equiv \ket{\{n_{I\bm{k}}\}} \otimes \ket{\psi_n}.\label{eq3a25}
\end{equation}
The spectral problem for the regularized free Hamiltonian $H_0^{\Lambda}$ in terms of these eigenstates is
\begin{equation}
    H_0^{\Lambda} \ket{\{n_{I\bm{k}}\};\psi_n} = \big(E_{B}^{\Lambda} + E_n^{\Lambda}\big) \ket{\{n_{I\bm{k}}\};\psi_n},
\end{equation}
with $E_{B}^{\Lambda}$ and $E_n^{\Lambda}$ given by (\ref{temp17}) and (\ref{temp16}), respectively. We want to compute the regularized energy shift $\Delta E_n^{\Lambda}$ of the electronic energy levels $E_{n}^{\Lambda}$
in the electromagnetic vacuum, resulting from interactions described by $H_{\text{int}}^{\Lambda}$, and we consider only bound states $\ket{\psi_n}$ in the Fermi sector. Since we will do so using perturbation theory at $\mathcal{O}(\alpha)$, it is useful to split up the interaction term
\begin{equation}
    H_{\text{int}}^{\Lambda} = H_{\text{int}(1)}^{\Lambda} + H_{\text{int}(2)}^{\Lambda},\label{eq3a28}
\end{equation}
where 
\begin{equation}
    H_{\text{int}(1)}^{\Lambda} = 2 \pi \int d\bm{x}\, \|\bm{p}_T(\bm{x})\|^2 - \frac{1}{2} \int d\bm{x}\, \bm{m}_{D}(\bm{x}) \cdot \bm{b}(\bm{x})\label{eq3a29}
\end{equation}
are the interactions that are already $\mathcal{O}(\alpha)$ and will be computed at the first order, while 
\begin{equation}
    H_{\text{int}(2)}^{\Lambda} = - \int d\bm{x}\, \bm{p}(\bm{x}) \cdot \bm{d}_T(\bm{x}) - \int d\bm{x}\, \bm{m}_{P}(\bm{x})\cdot\bm{b}(\bm{x})\label{eq3a30}
\end{equation}
are the interactions that are $\mathcal{O}(\sqrt{\alpha})$ and do not contribute to $H_{\text{int}(1)}^{\Lambda}$, since the vacuum expectation values of $\bm{d}_T(\bm{x})$ and $\bm{b}(\bm{x})$ vanish; these will be computed at the second order. We define
\begin{align}
    \ket{\Psi_{n}} &\equiv \ket{\text{vac}\,;\psi_n},\label{eq3c1} \\
    \ket{\Psi_{\delta}} &\equiv \ket{1_{I\bm{k}}\,;\psi_m}.\label{eq3c2}
\end{align}
In terms of these states, the regularized correction to the electronic energy levels $E_n^{\Lambda}$ is
\begin{equation}
    \Delta E_{n}^{\Lambda} = \Delta E_{n(1)}^{\Lambda} + \Delta E_{n(2)}^{\Lambda},\label{eq3a32}
\end{equation}
where
\begin{equation}
    \Delta E_{n(1)}^{\Lambda} = \bra{\Psi_{n}} H_{\text{int}(1)}^{\Lambda} \ket{\Psi_{n}}\label{eq3a33}
\end{equation}
is the \textit{first-order correction}, while 
\begin{equation}
    \Delta E_{n(2)}^{\Lambda} = \sum\limits_{\delta\neq n} \frac{\abs{\bra{\Psi_{\delta}}H_{\text{int}(2)}^{\Lambda}\ket{\Psi_{n}}}^2}{E_{n} - E_{\delta}}\label{eq3a34}
\end{equation}
is the \textit{second-order correction}.

\subsection{First-order correction}\label{subsection3C}
The first-order correction is
\begin{align}
    \Delta E_{n(1)}^{\Lambda} =  - \frac{1}{2} \int d\bm{x} \bra{\Psi_{n}}\bm{m}_D(\bm{x})\cdot\bm{b}(\bm{x}) \ket{\Psi_{n}}\nonumber \\
    + 2\pi \int d\bm{x} \bra{\Psi_{n}}\|\bm{p}_T(\bm{x})\|^2\ket{\Psi_{n}},\label{eq3c4}
\end{align}
where we understand that the right-hand-side is to be regularized with a cutoff at $\|\bm{k}\|= \Lambda$. The first term is the diamagnetic contribution to the regularized energy shift, which is more commonly written as
\begin{equation}
    + \frac{1}{2} \iint d\bm{x}d\bm{x}'\, \bra{\Psi_n}b^i(\bm{x}) O^{ij}(\bm{x},\bm{x}') b^j(\bm{x}')\ket{\Psi_n},\label{temporary1}
\end{equation}
where second-quantized \textit{diamagnetization field} is \cite{Healy}
\begin{equation}
    O^{ij}(\bm{x},\bm{x}') = \frac{e}{2mc^2} \int d\bm{y}\, \alpha^{ik}(\bm{x};\bm{y},\bm{R}) \alpha^{jk}(\bm{x}';\bm{y},\bm{R})\rho^e(\bm{y}).\label{temporary2}
\end{equation}
We show in Appendix \ref{appendixb} that in the continuum limit (\ref{cont}) the diamagnetic contribution to the first-order correction is given by
\begin{widetext}
\begin{align}
    -\frac{1}{2} \int d\bm{x}\, \bra{\Psi_{n}}\bm{m}_{D}(\bm{x})\cdot\bm{b}(\bm{x})\ket{\Psi_{n}} = \frac{\alpha}{2\pi} \left(\frac{\hbar^2 N_e}{m}\right) \Lambda^2 - \frac{\hbar}{mc} \int d\bm{x}\, \rho_{nn}^e(\bm{x}) \expval{\bm{a}(\bm{x})\cdot\bm{\nabla}\Phi(\bm{x},\bm{R})}_{\text{vac}}\nonumber \\
    + \frac{\hbar^2}{2me } \int d\bm{x}\, \rho_{nn}^e(\bm{x}) \expval{\|\bm{\nabla}\Phi(\bm{x},\bm{R})\|^2}_{\text{vac}}, \label{eq3c15}
\end{align}
\end{widetext}
where the expectation value is taken in the vacuum state of the radiation field. We will see below that when the full energy shift is calculated the latter two terms on the right-hand-side are cancelled by terms in the second-order correction $\Delta E_{n(2)}^{\Lambda}$.

Consider next the second term on the right-hand-side of (\ref{eq3c4}). Inserting a resolution of the identity in $\mathcal{H}_F$ with respect to the many-body eigenstates $\{\ket{\psi_m}\}_{m}$, the expectation value
\begin{align}
    \bra{\psi_n}p^i(\bm{x}') p^j(\bm{x})\ket{\psi_n} = \sum\limits_{m} p_{nm}^i(\bm{x}') p_{mn}^j(\bm{x}),\label{eq3c6}
\end{align}
where $p_{nm}^i(\bm{x}) \equiv \bra{\psi_n}p^i(\bm{x})\ket{\psi_m}$. Using the Fourier integral representation of the transverse delta function \cite{MolQED}
\begin{equation}
    \delta^{ij}_T(\bm{x}'-\bm{x}) = \int \frac{d\bm{k}}{(2\pi)^3} \delta_T^{ij}(\bm{k}) e^{i\bm{k}\cdot(\bm{x}'-\bm{x})},\label{eq3c7}
\end{equation}
(recall (\ref{tdelta})) we can (after regularization) write
\begin{align}
     2\pi \int d\bm{x}\, \|\bm{p}_T(\bm{x})\|^2 = \frac{1}{4\pi^2}\sum_m \int_{\|\bm{k}\|\leq \Lambda} d\bm{k}\, \delta_T^{ij}(\bm{k})\nonumber \\
    \times \iint d\bm{x}d\bm{x}'\, p_{nm}^i(\bm{x}') p_{mn}^j(\bm{x}) e^{i\bm{k}\cdot(\bm{x}'-\bm{x})}.\label{eq3c8}
\end{align}
To proceed, it is convenient to recast the angular part of the Fourier integral into an expression in coordinate space. Explicitly, we write \cite{Salam2}
\begin{align}
    \frac{1}{4\pi} \int d\Omega_{\bm{k}}\, &
    \delta_T^{ij}(\bm{k})e^{i\bm{k}\cdot(\bm{x}'-\bm{x})} = \frac{1}{\|\bm{k}\|^2} \tau^{ij}\big(k\|\bm{x}'-\bm{x}\|\big),\label{eq3c9}
\end{align}
where we have defined
\begin{equation}
    \tau^{ij}\big(k\|\bm{x}'-\bm{x}\|\big) \equiv \left(- \delta^{ij} \partial^2  + \partial^i \partial^j \right) F(\bm{x},\bm{x}';k),
\end{equation}
\\
We have introduced the abbreviated notation $d\Omega_{\bm{k}} = \sin\theta d\theta d\phi$ and $k = \|\bm{k}\|$, $\partial^i \equiv \partial/\partial x^i$ and $\partial^2 \equiv \nabla^2$, and have also defined
\begin{equation}
    F(\bm{x},\bm{x}';k) \equiv \frac{\sin(k\|\bm{x}-\bm{x}'\|)}{k\|\bm{x}-\bm{x}'\|}.\label{temp8}
\end{equation}
Relating the polarization field operators to the total charge density through (\ref{temp5}), the contribution to $\Delta E_{n(1)}^{\Lambda}$ coming from the transverse polarization fields is
\begin{widetext}
\begin{align}
    2\pi \int d\bm{x}\, \bra{\Psi_n}\|\bm{p}_T(\bm{x})\|^2\ket{\Psi_n} =  \frac{1}{\pi} \sum_{m} \int_0^{\Lambda} dk\, \iint d\bm{x} d\bm{x}'\, \tau^{ij}\big(k\|\bm{x}'-\bm{x}\|\big)\, p_{nm}^i(\bm{x}') p_{mn}^j(\bm{x}).\label{eq3c10}
\end{align}
Hence the total first-order correction (\ref{eq3c4}), the sum of (\ref{eq3c15}) and (\ref{eq3c10}), is
\begin{align}
    \Delta E_{n(1)}^{\Lambda} &= \frac{\alpha}{2\pi} \left(\frac{\hbar^2 N_e}{m}\right) \Lambda^2  + \frac{1}{\pi} \sum_{m} \int_0^{\Lambda} dk\, \iint d\bm{x} d\bm{x}'\, \tau^{ij}\big(k\|\bm{x}'-\bm{x}\|\big)\, p_{nm}^i(\bm{x}') p_{mn}^j(\bm{x})\nonumber \\
    &- \frac{\hbar}{mc} \int d\bm{x}\, \rho_{nn}^e(\bm{x}) \expval{\bm{a}(\bm{x})\cdot\bm{\nabla}\Phi(\bm{x},\bm{R})}_{\text{vac}} + \frac{\hbar^2}{2m e} \int d\bm{x}\, \rho_{nn}^e(\bm{x}) \expval{\|\bm{\nabla}\Phi(\bm{x},\bm{R})\|^2}_{\text{vac}}.\label{eq3c102}
\end{align}
\end{widetext}

\subsection{Second-order correction}\label{subsection3D}
To compute the second-order correction (\ref{eq3a34}), we begin with the matrix element
\begin{align}
    \bra{\Psi_{\delta}}H_{\text{int}(2)}^{\Lambda}\ket{\Psi_{n}} &= - \int d\bm{x}\, p_{mn}^j(\bm{x}) 
    \bra{1_{I\bm{k}}}d_T^j(\bm{x})\ket{\text{vac}}\nonumber \\
    &- \int d\bm{x}\, m_{P,mn}^j(\bm{x}) \bra{1_{I\bm{k}}}b^j(\bm{x})\ket{\text{vac}},\label{eq3d1}
\end{align}
where $m_{P,mn}^j(\bm{x}) \equiv \bra{\psi_m}m_P^j(\bm{x})\ket{\psi_n}$ denotes the \textit{paramagnetic} contribution to the magnetization. Inserting the mode expansions (\ref{eq3a13}), and using that the energy difference between the states $\ket{\Psi_{\delta}}$ and $\ket{\Psi_{n}}$ is $E_{\delta} - E_n = E_{mn} + \hbar \omega_{\bm{k}}$, where $E_{mn} \equiv E_m - E_n$, the regularized sum in expression (\ref{eq3a34}) is
\begin{equation}
    \sum_{\delta} \to \sum_{m}\sum_{I}\sum_{\|\bm{k}\|\leq \Lambda},\label{eq3d4}
\end{equation}
and after taking the continuum limit (\ref{cont}), we find
\begin{widetext}
\begin{align}
    \Delta E_{n(2)}^{\Lambda} = - \frac{\hbar c}{4\pi^2}\, \mathrm{p.v.} \sum_{m} \int_{\|\bm{k}\|\leq\Lambda} d\bm{k}\|\bm{k}\|\, \iint d\bm{x}d\bm{x}'\, \frac{e^{i\bm{k}\cdot(\bm{x}'-\bm{x})}}{E_{mn} + \hbar \omega_{\bm{k}}} \Bigg\{
    \delta_T^{ij}(\bm{k})\Big( p_{nm}^i(\bm{x}') p_{mn}^j(\bm{x})+ m_{P,nm}^i(\bm{x}') m_{P,mn}^j(\bm{x})\Big)\nonumber
    \\
    - \left(\varepsilon^{ijp}\frac{k^p}{\|\bm{k}\|}\right) \Big(p_{nm}^i(\bm{x}') m_{P,mn}^j(\bm{x}) + m_{P,nm}^i(\bm{x}') p_{mn}^j(\bm{x})  \Big)\Bigg\},\label{eq3d5}
\end{align}
where $\mathrm{p.v.}(\cdot)$ denotes the Cauchy principal value, which must be included for states $\ket{\psi_n}$ with $n \neq 0$ since ``downward" resonant transitions for which $E_{nm} = \hbar\omega_{\bm{k}}$ are energetically permitted \cite{dispersionv2}. The simplification of this expression is largely formulaic and has been relegated to Appendix \ref{appendixc}. There we show that it can be written as
    \begin{align}
         \Delta E_{n(2)}^{\Lambda} = &- \frac{1}{\pi}\, \mathrm{p.v.} \sum_{m} \int_0^{\Lambda} dk \frac{\hbar c k}{E_{mn} + \hbar c k}\, \iint d\bm{x}d\bm{x}'\, \tau^{ij}\big(k\|\bm{x}'-\bm{x}\|\big)\, \Big(p_{nm}^i(\bm{x}') p_{mn}^j(\bm{x}) + m_{P,nm}^i(\bm{x}') m_{P,mn}^j(\bm{x})\Big)\nonumber
    \\
     &- \frac{i}{\pi}\, \mathrm{p.v.} \sum_{m} \int_0^{\Lambda} dk \frac{\hbar c k^2}{E_{mn} + \hbar c k}\, \iint d\bm{x}d\bm{x}'\, \sigma^{ij}\big(k\|\bm{x}'-\bm{x}\|\big)\, \Big(p_{nm}^i(\bm{x}') m_{P,mn}^j(\bm{x}) + m_{P,nm}^i(\bm{x}') p_{mn}^j(\bm{x})\Big),\label{eq3d7}
    \end{align}
where 
\begin{equation}
    \sigma^{ij}\big(k\|\bm{x}'-\bm{x}\|\big) \equiv \frac{i}{4\pi} \int d\Omega_{\bm{k}}\,\varepsilon^{ipj} k^p e^{i \bm{k}\cdot(\bm{x}'-\bm{x})}.\label{eq3d6}
\end{equation}
Further simplifications can be made using the relation (\ref{temp5}) and the Schrödinger-picture identity
\begin{equation}
    \bm{j}_{P,nm}(\bm{x}) = (i\hbar)^{-1} E_{mn}^R \bm{p}_{nm}(\bm{x}) + c \bm{\nabla} \times \bm{m}_{P,nm}(\bm{x}),\label{paramag}
\end{equation}
\\
proven in Appendix \ref{appendixc1}. After a tedious but straightforward calculation, the total second-order correction is
\begin{align}
    \Delta E_{n(2)}^{\Lambda} = - \frac{\hbar}{4 \pi^2 c}\, \mathrm{p.v.} \sum_{m} \int_{\|\bm{k}\|\leq\Lambda} d\bm{k}\, \frac{\|\bm{k}\|^{-1}}{E_{mn} + \hbar \omega_{\bm{k}}} \iint d\bm{x}d\bm{x}'\, j_{P,nm}^i(\bm{x}') j_{P,mn}^j(\bm{x}) 
    \delta_T^{ij}(\bm{k})e^{i\bm{k}\cdot(\bm{x}'-\bm{x})}\nonumber \\
    + \frac{\hbar}{mc} \int d\bm{x}\, \rho_{nn}^e(\bm{x}) \expval{\bm{a}(\bm{x})\cdot\bm{\nabla}\Phi(\bm{x},\bm{R})}_{\text{vac}} - \frac{\hbar^2}{2m e} \int d\bm{x}\, \rho_{nn}^e(\bm{x}) \expval{\|\bm{\nabla}\Phi(\bm{x},\bm{R})\|^2}_{\text{vac}}\nonumber \\
    - \frac{1}{\pi} \sum_{m} \int_0^{\Lambda} dk\, \iint d\bm{x} d\bm{x}'\, \tau^{ij}\big(k\|\bm{x}'-\bm{x}\|\big)\, p_{nm}^i(\bm{x}') p_{mn}^j(\bm{x}).\label{secondorder}
\end{align}
Hence, combining the first-order (\ref{eq3c102}) and second-order (\ref{secondorder}) corrections, the regularized energy shift is
\begin{equation}
        \Delta E_{n}^{\Lambda} = \frac{\alpha}{2\pi} \left(\frac{\hbar^2 N_e}{m}\right) \Lambda^2 - \frac{\hbar}{4 \pi^2 c}\, \mathrm{p.v.} \sum_{m} \int_{\|\bm{k}\|\leq\Lambda} d\bm{k}\, \frac{\|\bm{k}\|^{-1}}{E_{mn} + \hbar \omega_{\bm{k}}} \iint d\bm{x}d\bm{x}'\, j_{P,nm}^i(\bm{x}') j_{P,mn}^j(\bm{x}) 
       \delta_T^{ij}(\bm{k})e^{i\bm{k}\cdot(\bm{x}'-\bm{x})}.\label{eq3d17}
    \end{equation}
We can also write $\Delta E_{n}^{\Lambda}$ in terms of the polarization vectors $\{\bm{e}_{I}(\bm{k})\}_I$:
    \begin{equation}
        \Delta E_{n}^{\Lambda} = \frac{\alpha}{2\pi} \left(\frac{\hbar^2 N_e}{m}\right) \Lambda^2 - \frac{\hbar}{4 \pi^2 c}\, \mathrm{p.v.} \sum_{m} \int_{\|\bm{k}\|\leq\Lambda} d\bm{k}\, \frac{\|\bm{k}\|^{-1}}{E_{mn} + \hbar \omega_{\bm{k}}} \sum_{I} \abs{\bm{e}_{I}(\bm{k})\cdot\tilde{\bm{j}}_{P,mn}(\bm{k})}^2,\label{eq3d18}
    \end{equation}
\end{widetext}
where $\tilde{\bm{j}}_P(\bm{k})$ is the Fourier transform 
\begin{equation}
    \tilde{\bm{j}}_P(\pm\bm{k}) = \int d\bm{x}\, \bm{j}_P(\bm{x}) e^{\mp i \bm{k}\cdot\bm{x}}.
\end{equation}
Existing treatments \cite{Bethe, Au, Grotch} differ from ours either because they are computed within a first-quantized framework, or because they do not include spatial variations of the electromagnetic field over the molecule. 
If those variations would be neglected, the polarization sum in (\ref{eq3d18}) would reduce to
\begin{equation}
    \sum_{I} \abs{\bm{e}_{I}(\bm{k})\cdot\bm{\mathrm{P}}_{nm}}^2,
\end{equation}
\\
where 
\begin{equation}
    \bm{\mathrm{P}} = \frac{\hbar}{i} \int d\bm{x}\, \psi^{\dagger}(\bm{x})\bm{\nabla}\psi(\bm{x})\label{temp31}
\end{equation}
is the total (second-quantized) momentum operator. This distinction highlights the difference between our approach and existing treatments, which generally only apply to hydrogenic atoms. If one is interested in many-electron atoms, molecules, or more general charge-current distributions, then it is necessary to couple the radiation field to the current \textit{density} rather than the total momentum. In doing so, our result becomes valid \textit{to all orders} in the multipole expansion.

\section{Renormalization}\label{section4}
\subsection{Renormalization of $H_{0}^{\Lambda}$}
We first renormalize the free Hamiltonian $H_0^{\Lambda}$, which produces the vacuum divergence $E_{0,B}^{\Lambda}$ of free quantum electrodynamics given in (\ref{eq3T1}) and the divergence $E_{0,F}^{\Lambda}$ computed in Appendix \ref{appendixA}. Since both of these divergences are \textit{static}, we can trivially remove them from $H_0^{\Lambda}$ by addition of static counterterms $\delta H_{B}(\Lambda)$ and $\delta H_F(\Lambda)$ to $H_{B}^{\Lambda}$ and $H_{F}^{\Lambda}$. The renormalized free Hamiltonian is
\begin{equation}
    H_0^{R} = H_B^R \otimes \mathbb{I}_F + \mathbb{I}_B \otimes H_F^R, 
\end{equation}
where the renormalized Maxwell and electronic terms are
\begin{align}
    H_{B}^{R} &= H_{B}^{\Lambda} + \delta H_{B}(\Lambda), \\
    H_{F}^{R} &= H_{F}^{\Lambda} + \delta H_{F}(\Lambda),
\end{align}
and the counterterms (in minimal subtraction \cite{Schwartz}) are
\begin{align}
    \delta H_{B}(\Lambda) &= - \left[\frac{\hbar c V}{8\pi^2} \Lambda^4\right] \mathbb{I}_B,\label{ct1} \\
    \delta H_{F}(\Lambda) &= - \left[\frac{1}{\pi} \left(e^2 N_e + 2 e N_e \sum_{N} q_N  + \sum_N q_N^2\right)\Lambda\right] \mathbb{I}_F.\label{ct2}
\end{align}
The energy eigenvalues computed from $H_{B}^R$ and $H_{F}^R$ are finite in the limit $\Lambda \to \infty$. In particular, the renormalized Maxwell term is
\begin{equation}
    H_B^R = \sum_{I\bm{k}} \hbar \omega_{\bm{k}} a_{I\bm{k}}^{\dagger} a_{I\bm{k}},
\end{equation}
while the renormalized electronic term is
\begin{equation}
    H_F^R = T + U^R,
\end{equation}
where the kinetic term is
\begin{equation}
    T = \frac{\hbar^2}{2m} \int d\bm{x}\, \bm{\nabla}\psi^{\dagger}(\bm{x})\cdot\bm{\nabla}\psi(\bm{x})\label{H0}
\end{equation}
and the renormalized electrostatic interaction term is
\begin{align}
    U^R = \frac{1}{2} \iint d\bm{x}d\bm{x}'\, \psi^{\dagger}(\bm{x}) \psi^{\dagger}(\bm{x}')\left(\frac{e^2}{\|\bm{x}-\bm{x}'\|}\right) \psi(\bm{x}') \psi(\bm{x})\nonumber \\
    + \int d\bm{x}\, \psi^{\dagger}(\bm{x}) \mathrm{V}(\bm{x}) \psi(\bm{x}).\label{eqTEMP11}
\end{align}
The renormalized electronic energy levels are the expectation values $E_{n}^R = \expval{H_F^R}_n$. There are a host of techniques in atomic and molecular physics, and in condensed matter physics, that have been designed to compute expectation values of the operator given in (\ref{eqTEMP11}) \cite{Cohen}.

\subsection{Renormalization of $H_{\text{int}}^{\Lambda}$}
The renormalized interaction term is defined by
\begin{equation}
    H_{\text{int}}^{R} = H_{\text{int}}^{\Lambda} + \delta H_{\text{int}}(\Lambda),\label{temp32}
\end{equation}
where $\delta H_{\text{int}}(\Lambda)$ will be determined below. As outlined in Section \ref{section2}, our renormalization scheme is based on the observation that the $\Lambda$-dependent contribution to the regularized energy shift (\ref{eq3d18}) coming from free electrons, described by the Hamiltonian $H_F^{(0)} \equiv T$ (see (\ref{H0})), is unobservable and should be subtracted from $\Delta E_{n}^{\Lambda}$. What remains will be the observable correction to the \textit{bound state} electronic energy levels.

Thus we must first determine the contribution to $\Delta E_{n}^{\Lambda}$ coming from free electrons, described by $H_{F}^{(0)}$; denote by $\{\ket{\varphi_{n'}}\}_{n'}$ the set of many-body eigenstates thereof. Because this Hamiltonian commutes with the total momentum operator $\bm{\mathrm{P}}$, we can choose each energy eigenstate $\ket{\varphi_{n'}}$ to be an eigenstate of $\bm{\mathrm{P}}$. Then for an arbitrary many-body eigenstate $\ket{\varphi_{n'}}$ we have
\begin{align}
    \Delta E_{\text{free}}^{\Lambda} &\equiv \bra{\varphi_{n'}}H_{\text{int}}^{\Lambda}\ket{\varphi_{n'}}, \\
    \Delta E_{\text{free}}^{R} &\equiv \bra{\varphi_{n'}}H_{\text{int}}^{R}\ket{\varphi_{n'}}.\label{temp34}
\end{align}
We choose the 
counterterms in $\delta H_{\text{int}}(\Lambda)$ to cancel the $\Lambda$-dependent terms in $\Delta E_{\text{free}}^{\Lambda}$, so that the renormalized energy shift $\Delta E_{\text{free}}^{R}$ vanishes for an arbitrary state $\ket{\varphi_{n'}}$, \textit{i.e.}, an arbitrary distribution of electron momenta.

To compute $\Delta E_{\text{free}}^{\Lambda}$, it is useful to expand the electron field operators in a basis of plane waves
\begin{equation}
    \psi(\bm{x}) = \frac{1}{\sqrt{V}} \sum_{\bm{q}} e^{i\bm{q}\cdot\bm{x}} b_{\bm{q}},
\end{equation}
where we return to considering a system in volume $V$, and where the anticommutation relations (\ref{eq2a10}) imply that the electronic creation and annihilation operators satisfy the canonical anticommutation relations
\begin{align}
    \big[b_{\bm{q}}, b_{\bm{q}'}^{\dagger}\big]_+ &= \delta_{\bm{q}\bm{q}'},\nonumber \\
    \big[b_{\bm{q}}, b_{\bm{q}'}\big]_+ &= \big[b_{\bm{q}}^{\dagger}, b_{\bm{q}'}^{\dagger}\big]_+ = 0.
\end{align}
In this plane-wave expansion, the Fourier transform of the paramagnetic current density operator is
\begin{equation}
    \tilde{j}_P^j(\bm{k}) = \frac{\hbar e}{2m} \sum_{\bm{q}} \Big(2q^j - k^j\Big) b_{\bm{q}-\bm{k}}^{\dagger} b_{\bm{q}}.\label{eqTempT3}
\end{equation}
We obtain $\Delta E_{\text{free}}^{\Lambda}$ by replacing the many-body eigenstates of $H_F$ involved in (\ref{eq3d18}) with those of $H_F^{(0)}$, and replacing $E_{mn}$ with $E_{m'n'}^{(0)}$, where $E_{n'}^{(0)} = \bra{\varphi_{n'}}H_F^{(0)}\ket{\varphi_{n'}}$. Thus, for free electrons the regularized energy shift is
\begin{widetext}
\begin{equation}
    \Delta E_{\text{free}}^{\Lambda} = \frac{\alpha}{2\pi} \left(\frac{\hbar^2 N_e}{m}\right) \Lambda^2 - \frac{\hbar}{4\pi^2 c}\, \mathrm{p.v.} \sum\limits_{m'}\int_{\|\bm{k}\|\leq \Lambda} d\bm{k}\, \frac{\|\bm{k}\|^{-1}}{E_{m'n'}^{(0)} + \hbar \omega_{\bm{k}}} \sum_{I} \abs{\bm{e}_{I}(\bm{k})\cdot \tilde{\bm{j}}_{P,m'n'}(\bm{k})}^2.\label{eq146}
\end{equation}
\end{widetext}
To simplify the second term, consider the expression
\begin{align}
    \sum_{m'} \tilde{j}_{P,n'm'}^i(-\bm{k})\left(\frac{1}{E_{m'n'}^{(0)} + \hbar \omega_{\bm{k}}}\right)\tilde{j}_{P,m'n'}^j(\bm{k}),\label{eq5T1}
\end{align}
which features in the integrand of (\ref{eq146}). To proceed, 
we write the term in parentheses as
\begin{equation}
    \frac{1}{E_{m'n'}^{(0)}+ \hbar \omega_{\bm{k}} +i\delta} = -i \int_0^{\infty} ds\, e^{is \left(E_{m'}^{(0)} - E_{n'}^{(0)} + \hbar \omega_{\bm{k}} +i\delta\right)},\label{rep}
\end{equation}
\\
where we have introduced a convergence factor $i\delta$ (and the limit $\delta \to 0^+$ is understood), with which (\ref{eq5T1}) is
\begin{widetext}
\begin{equation}
    \sum_{m'} \tilde{j}_{P,n'm'}^i(-\bm{k})\left(\frac{1}{E_{m'n'}^{(0)} + \hbar \omega_{\bm{k}} + i\delta}\right)\tilde{j}_{P,m'n'}^j(\bm{k}) = -i\sum_{m'} \int_0^{\infty} ds\, e^{is(\hbar \omega_{\bm{k}} + i\delta)} \tilde{j}_{P,n'm'}^i(-\bm{k}) \expval{e^{is H_F^{(0)}} \tilde{j}_P^j(\bm{k}) e^{-isH_F^{(0)}}}_{m'n'},
\end{equation}
\\
where the subscript $m'n'$ indicates the matrix element between state $\ket{\varphi_{m'}}$ and state $\ket{\varphi_{n'}}$. The operator in the angular brackets on the right-hand-side is simply the paramagnetic current density operator in the interaction picture with $s = t/\hbar$ (which acts trivially on $\mathcal{H}_B$), and so we have
\begin{equation}
    e^{is H_F^{(0)}} \tilde{j}_P^j(\bm{k}) e^{-isH_F^{(0)}} = \frac{\hbar e}{2m} \sum_{\bm{q}} \Big(2q^j - k^j\Big)\, b_{\bm{q}-\bm{k}}^{\dagger} b_{\bm{q}} e^{is(E_{\hbar\omega_{\bm{k}}} - \hbar \bm{k}\cdot\bm{q}/m)},\label{intpicj}
\end{equation}
where $E_{\hbar\omega_{\bm{k}}} = \hbar^2 \omega_{\bm{k}}^2 / 2mc^2$. Therefore, using the identity (\ref{rep}) in reverse (and taking $\delta \to 0^+$) 
\begin{align}
    \sum_{m'} \tilde{j}_{P,n'm'}^i(-\bm{k}) \left(\frac{1}{E_{m'n'}^{(0)} + \hbar \omega_{\bm{k}}}\right) \tilde{j}_{P,m'n'}^j(\bm{k}) = \frac{\hbar^2 e^2}{4m^2} \sum_{\bm{q}\bm{q}'} \Big(2q'^i - k^i\Big) \Big(2q^j - k^j\Big) \left(\frac{1}{E_{\hbar\omega_{\bm{k}}} + \hbar \omega_{\bm{k}} - \hbar \bm{k}\cdot\bm{q}/m}\right)\nonumber \\
    \times \sum_{m'} \bra{\varphi_{n'}} b_{\bm{q}'}^{\dagger} b_{\bm{q}'-\bm{k}}\ketbra{\varphi_{m'}}{\varphi_{m'}} b_{\bm{q}-\bm{k}}^{\dagger}b_{\bm{q}} \ket{\varphi_{n'}}.\label{eqTemp10}
\end{align}
Since we are working with electrons that are non-relativistic, $\|\bm{q}\|\ll mc$, and we can drop the ``$\hbar \bm{k}\cdot\bm{q}/m$" term in the denominator. Recollecting the Fourier transforms of the paramagnetic current density operators using (\ref{eqTempT3}) we find
\begin{align}
    \sum_{m'} \tilde{j}_{P,n'm'}^i(-\bm{k}) \left(\frac{1}{E_{m'n'}^{(0)} + \hbar \omega_{\bm{k}}}\right) \tilde{j}_{P,m'n'}^j(\bm{k}) = \sum_{m'} \tilde{j}_{P,n'm'}^i(-\bm{k})\left(\frac{1}{E_{\hbar\omega_{\bm{k}}} + \hbar \omega_{\bm{k}}}\right)\tilde{j}_{P,m'n'}^j(\bm{k}),
\end{align}
from which follows
\begin{equation}
    \Delta E_{\text{free}}^{\Lambda} = \frac{\alpha}{2\pi} \left(\frac{\hbar^2 N_e}{m}\right) \Lambda^2\, - \frac{\hbar}{4\pi^2 c} \sum_{m'} \int_{\|\bm{k}\|\leq \Lambda} d\bm{k}\, \frac{\|\bm{k}\|^{-1}}{E_{\hbar\omega_{\bm{k}}} + \hbar \omega_{\bm{k}}} \sum_{I} \abs{\bm{e}_{I}(\bm{k}) \cdot \tilde{\bm{j}}_{P,m'n'}(\bm{k})}^2,
    \label{free-reg}
\end{equation}
\end{widetext}
where we have dropped the principal value notation, since the integrand has no poles (or infrared divergences). 

As discussed in Section \ref{section2}, we add to the regularized interaction term a counterterm $\delta H_{\text{int}}(\Lambda)$ whose $\Lambda$-dependence is chosen so that the renormalized interaction term (\ref{temp32})
is finite in the limit $\Lambda \to \infty$. However, instead of setting 
 the counterterm directly, we proceed by enforcing the condition that $\Delta E_{\text{free}}^{R}$ 
vanishes. Enforcing this condition requires that $\expval{\delta H_{\text{int}}(\Lambda)}_{n'}$ is the negative of the right-hand-side of (\ref{free-reg}) for all $\ket{\varphi_{n'}}$, and this can be satisfied by taking 
\begin{widetext}
\begin{equation}
    \delta H_{\text{int}}(\Lambda) = - \left[\frac{\alpha}{2\pi} \left(\frac{\hbar^2 N_e}{m}\right) \Lambda^2\right] \mathbb{I}_B \otimes \mathbb{I}_F\, + \frac{\hbar}{4\pi^2 c} \int_{\|\bm{k}\|\leq \Lambda} d\bm{k}\, \frac{\|\bm{k}\|^{-1}}{E_{\hbar\omega_{\bm{k}}} + \hbar \omega_{\bm{k}}}\sum_I \Big(\bm{e}_I(-\bm{k})\cdot\tilde{\bm{j}}_{P}^{\dagger}(\bm{k})\Big) \Big(\bm{e}_I(\bm{k})\cdot\tilde{\bm{j}}_{P}(\bm{k})\Big).\label{ct3}
\end{equation}
\end{widetext}

We have now collected all of the counterterms necessary to obtain the full renormalized Hamiltonian from
\begin{equation}
    H^{\Lambda} = H_B^{\Lambda}\otimes \mathbb{I}_F + \mathbb{I}_B \otimes H_F^{\Lambda} + H_{\text{int}}^{\Lambda}.
\end{equation}
The renormalized Hamiltonian is
\begin{equation}
    H^{R} = H^{\Lambda} + \delta H(\Lambda),
\end{equation}
where $H^{\Lambda}$ is given by (\ref{eq3a26}) and
\begin{equation}
    \delta H(\Lambda) = \delta H_B(\Lambda)\otimes \mathbb{I}_F + \mathbb{I}_B \otimes \delta H_F(\Lambda) + \delta H_{\text{int}}(\Lambda).
\end{equation}
Using (\ref{ct1},\ref{ct2}) and (\ref{ct3}), the counterterm $\delta H(\Lambda)$ is
\begin{widetext}
\begin{align}
    \delta H(\Lambda) = - \left[\frac{\alpha}{2\pi} \left(\frac{\hbar^2 N_e}{m}\right) \Lambda^2 + \frac{\hbar c V}{8\pi^2} \Lambda^4 + \frac{1}{\pi} \left(e^2 N_e + 2 e N_e \sum_{N} q_N +\sum_N q_N^2\right)\Lambda\right] \mathbb{I}_B \otimes \mathbb{I}_F\nonumber \\
    + \frac{\hbar}{4\pi^2 c} \int_{\|\bm{k}\|\leq \Lambda} d\bm{k}\, \frac{\|\bm{k}\|^{-1}}{E_{\hbar\omega_{\bm{k}}} + \hbar \omega_{\bm{k}}} \sum_I \Big(\bm{e}_I(-\bm{k})\cdot\tilde{\bm{j}}_{P}^{\dagger}(\bm{k})\Big) \Big(\bm{e}_I(\bm{k})\cdot\tilde{\bm{j}}_{P}(\bm{k})\Big).
\end{align}

\subsection{Renormalized energy shift}
We are now in a position to calculate the renormalized energy shift
\begin{equation}
    \Delta E_n^R = \bra{\text{vac}\,;\psi_n}H_{\text{int}}^R\ket{\text{vac}\,;\psi_n},
\end{equation}
where $\ket{\psi_n}$ is a (bound) eigenstate of the renormalized electronic Hamiltonian $H_F^R \ket{\psi_n} = E_n^R \ket{\psi_n}$ with eigenvalue $E_n^R$ given by the expectation value of (\ref{eqTEMP11}). The first counterterm in $\expval{\delta H_{\text{int}}(\Lambda)}_{n} \equiv \bra{\psi_n}\delta H_{\text{int}}(\Lambda)\ket{\psi_n}$ (\textit{cf}. the first term in (\ref{ct3})) trivially cancels the static self-energy in $\Delta E_n^{\Lambda}$ and we are left with
\begin{align}
    \Delta E_n^R = &- \frac{\hbar}{4 \pi^2 c}\, \mathrm{p.v.} \sum_{m} \int_{\|\bm{k}\|\leq \Lambda} d\bm{k}\, \frac{\|\bm{k}\|^{-1}}{E_{mn}^R + \hbar \omega_{\bm{k}}} \sum_{I} \abs{\bm{e}_{I}(\bm{k}) \cdot \tilde{\bm{j}}_{P,mn}(\bm{k})}^2\nonumber \\
    &+ \frac{\hbar}{4\pi^2 c} \int_{\|\bm{k}\|\leq \Lambda} d\bm{k}\, \frac{\|\bm{k}\|^{-1}}{E_{\hbar\omega_{\bm{k}}} + \hbar \omega_{\bm{k}}} \sum_{I} \expval{\Big(\bm{e}_I(-\bm{k})\cdot\tilde{\bm{j}}_{P}^{\dagger}(\bm{k})\Big) \Big(\bm{e}_I(\bm{k})\cdot\tilde{\bm{j}}_{P}(\bm{k})\Big)}_n,
\end{align}
the second line being the second counterterm in (\ref{ct3}). Introducing a resolution of the identity in $\mathcal{H}_F$ with respect to the states $\{\ket{\psi_m}\}_m$, we can collect these terms together
\begin{align}
    \Delta E_n^R = - \frac{\hbar}{4 \pi^2 c}\, \mathrm{p.v.} \sum_{m} \int_{\|\bm{k}\|\leq \Lambda} d\bm{k}\, \left(\frac{\|\bm{k}\|^{-1}}{E_{mn}^R + \hbar \omega_{\bm{k}}} -\frac{\|\bm{k}\|^{-1}}{E_{\hbar\omega_{\bm{k}}} + \hbar \omega_{\bm{k}}}\right) \sum_{I} \abs{\bm{e}_{I}(\bm{k}) \cdot \tilde{\bm{j}}_{P,mn}(\bm{k})}^2.\label{temp28}
\end{align}
To simplify the expression in the parentheses, we use the following identity, which we prove in 
Appendix \ref{appendixe}:
\begin{equation}
    \sum_{m} \tilde{j}_{P,nm}^i(-\bm{k}) \frac{1}{E_{mn}^R + \hbar \omega_{\bm{k}}} \tilde{j}_{P,mn}^j(\bm{k}) = \sum_{m} \left(\frac{\tilde{j}_{P,nm}^i(-\bm{k}) \tilde{j}_{P,mn}^j(\bm{k})}{E_{\hbar\omega_{\bm{k}}} + \hbar \omega_{\bm{k}}} -  \frac{\tilde{j}_{P,nm}^i(-\bm{k}) \big\langle\big[U^R, \tilde{j}_P^j(\bm{k})\big]_-\big\rangle_{mn}}{(E_{\hbar\omega_{\bm{k}}} + \hbar \omega_{\bm{k}})(E_{mn}^R + \hbar \omega_{\bm{k}})}\right).\label{eqTEMP14}
\end{equation}
The first term on the right of the above equality exactly cancels the second term in the parentheses of (\ref{temp28}), and so the renormalized energy shift is
\begin{equation}
    \Delta E_{n}^{R} = \frac{\hbar}{4\pi^2 c}\, \mathrm{p.v.} \sum_{m} \int d\bm{k}\|\bm{k}\|^{-1}\, \frac{\delta_T^{ij}(\bm{k})\,}{(E_{\hbar\omega_{\bm{k}}} + \hbar \omega_{\bm{k}})(E_{mn}^R + \hbar \omega_{\bm{k}})}\Big(\tilde{j}_{P,nm}^i(-\bm{k})\big\langle\big[U^R, \tilde{j}_P^j(\bm{k})\big]_-\big\rangle_{mn}\Big),\label{eq5T5}
\end{equation}
\end{widetext}
where we have removed the regulator, since this expression is finite in the limit $\Lambda\to\infty$.

Before returning to the multipole formalism to study contributions to $\Delta E_n^{R}$ order-by-order in the multipole expansions, we first quote the result that one obtains by directly calculating this expression in terms of the paramagnetic current density; the details can be found in Appendix \ref{appendixg}. Replacing $E_{mn}^R$ by a ``reference" or average value $\bar{E}_n^R$ \cite{Bethe, MolQED}, the result of this calculation is
\begin{equation}
    \Delta E_{n}^R = - \frac{4 \alpha \hbar^2}{3 m^2 c^2}\sum_N q_N \frac{\expval{\rho^e(\bm{R}+\bm{d}_N)}_n}{1 - \bar{E}_n^R / 2mc^2} \ln(\frac{2m c^2}{\abs{\bar{E}_n^R}}).
    \label{temp210}
\end{equation}
To compare with existing results, we consider the case of a single hydrogenic atom at $\bm{R}$. We can thereby replace $q_N \expval{\rho^e(\bm{R}+\bm{d}_N)}_n$ with $- e^2 Z \abs{\phi_{n00}(\bm{R})}^2$, where $\phi_{n\ell m}(\bm{x})$ are the usual hydrogen wavefunctions. Moreover, using $\abs{\phi_{n00}(\bm{R})}^2 = Z^3/n^3 \pi a_0^3$ we have \cite{Ford1,Au}
\begin{equation}
    \Delta E_{n}^R = \frac{4 \alpha^5 Z^4}{3 \pi n^3} \left(\frac{mc^2}{1 - \bar{E}_n^R / 2mc^2}\right) \ln(\frac{2 m c^2}{\abs{\bar{E}_n^R}}).
\end{equation}
For a hydrogen atom ($Z=1$) we take the average excitation energy for the $2s$ level to be $\abs{\bar{E}_{2s}^R} = 16.64\; \text{Ryd}$, from which we find the energy shift to be $\Delta E_{2s}^R \approx 1051\; \text{MHz}$, which is in good agreement with the experimental value of $1054\; \text{MHz}$ \cite{Lamb, MolQED}. To make the electric dipole approximation, we replace the denominator with unity and thereby obtain the standard result \cite{Welton, MolQED, Bethe}
\begin{equation}
    \Delta E_{n}^R = \frac{4 \alpha^5}{3 \pi} mc^2 \ln(\frac{2m c^2}{\abs{\bar{E}_n^R}}).\label{eq5T11}
\end{equation}

\subsection{Multipole expansion}
A primary advantage of using the multipole formalism over minimal coupling is the ability to perform localized multipole expansions of the polarization and magnetization fields. While renormalization is better done with the minimal coupling form of $\Delta E_n^R$ as above, we now rewrite it in terms of these polarization and magnetization fields. As we discuss below, the multipole expansions are only valid for small $\|\bm{k}\|$, and in particular for $\|\bm{k}\| < \Lambda$. This means that we can drop the factor of $E_{\hbar\omega_{\bm{k}}}$ in the denominator of (\ref{temp28}), and use a modified form of the identity in the parentheses of (\ref{eqTEMP14}), namely
\begin{equation}
    \frac{1}{E_{mn}^R + \hbar\omega_{\bm{k}}} - \frac{1}{\hbar\omega_{\bm{k}}} = -\frac{E_{mn}^R}{(\hbar\omega_{\bm{k}})(E_{mn}^R + \hbar\omega_{\bm{k}})},
\end{equation}
to write the renormalized energy shift as
\begin{widetext}
\begin{equation}
    \Delta E_n^R = \frac{1}{4\pi^2 c^2}\,\mathrm{p.v.} \sum_m \int d\bm{k} \|\bm{k}\|^{-2} \frac{E_{mn}^R}{E_{mn}^R + \hbar\omega_{\bm{k}}} \sum_I\abs{\bm{e}_I(\bm{k})\cdot\tilde{\bm{j}}_{P,mn}(\bm{k})}^2.
\end{equation}
Then to rewrite $\Delta E_n^R$ in terms of the polarization and magnetization fields, we use the identity (\ref{paramag}) to write the renormalized energy shift as
\begin{align}
    \Delta E_{n}^R &= \frac{1}{4\pi^2}\,\mathrm{p.v.} \sum_m \int d\bm{k} \left(\frac{E_{mn}^R}{E_{mn}^R + \hbar \omega_{\bm{k}}}\right)\sum_I\left[\left(\frac{E_{mn}^R}{\hbar \omega_{\bm{k}}}\right)^2 \Big|\bm{e}_I(\bm{k})\cdot\tilde{\bm{p}}_{mn}(\bm{k})\Big|^2 +\Big|\bm{e}_I(\bm{k})\cdot\tilde{\bm{m}}_{P,mn}(\bm{k})\Big|^2 \right]\nonumber \\
    &+ \frac{1}{2\pi^2}\, \mathrm{p.v.} \sum_m \int d\bm{k} \left(\frac{E_{mn}^R}{E_{mn}^R + \hbar \omega_{\bm{k}}}\right) \left(\frac{E_{mn}^R}{\hbar \omega_{\bm{k}}}\right)\sum_I s_I \Im\left[\Big(\bm{e}_I(-\bm{k})\cdot\tilde{\bm{p}}_{nm}(-\bm{k})\Big)\Big(\bm{e}_I(\bm{k})\cdot\tilde{\bm{m}}_{P,mn}(\bm{k})\Big)\right]
    .\label{eq5T9}
\end{align}
\end{widetext}
This is our general expression for the renormalized energy shift in the multipole formalism. As we discuss at the end of this section, making the electric dipole approximation for the ground state $n = 0$ of atomic hydrogen yields the usual Bethe result \cite{Bethe}. However, our result above holds for ground and excited states of a general localized charge-current distribution, and includes contributions from the full electric and magnetic multipole series. And it accounts for spatial variations in the electromagnetic field, which are usually neglected by replacing the polarization and magnetization fields with the electric and magnetic dipole moments. Of course, in dropping the factors of $E_{\hbar\omega_{\bm{k}}}$ above we have reintroduced the ``Bethe log" divergences, but in any case the multipole expansions introduce stronger ultraviolet divergences that must be regularized, as discussed below. 

To obtain the multipole expansions, we take as the defining curve for the ``relators" $s^i(\bm{x};\bm{y},\bm{R})$ and $\alpha^{ij}(\bm{x};\bm{y},\bm{R})$ a straight-line path
\begin{equation}
    \bm{z}(\lambda) = \bm{R} + \lambda(\bm{y}-\bm{R})
\end{equation}
\\
with unit parametrization $\lambda \in [0,1]$, in which case \cite{Healy}
\begin{align}
    s^i(\bm{x};\bm{y},\bm{R}) &= (y^i - R^i) \int_0^1 d\lambda \delta (\bm{x} - \bm{R} - \lambda(\bm{y} - \bm{R})),\nonumber \\
    \alpha^{ij}(\bm{x};\bm{y},\bm{R}) &= \varepsilon^{ijk} (y^k - R^k) \int_0^1 d\lambda \lambda \delta(\bm{x} - \bm{R} - \lambda(\bm{y} - \bm{R})).
\end{align}
We then expand these expressions in a formal Taylor series with appropriate radius of convergence. Inserting the result back into the polarization and magnetization fields, the full electric and paramagnetic multipole expansions are then given by
\begin{align}
    p^i(\bm{x}) &= \sum_{n=0}^{\infty} (-1)^n \mu^{i_1 \dots i_n i} \partial^{i_1} \dots \partial^{i_n} \delta(\bm{x}-\bm{R}),\nonumber \\
    m_P^i(\bm{x}) &= \sum_{n=0}^{\infty} (-1)^n \nu^{i_1\dots i_n i} \partial^{i_1}\dots \partial^{i_n}\delta(\bm{x}-\bm{R}),\label{fullmult}
\end{align}
where the $n^{\text{th}}$-order electric and paramagnetic multipole moment operators are
\begin{widetext}
\begin{align}
    \mu^{i_1 \dots i_n i} &= \frac{1}{n!} \int d\bm{r}\, \big(r^{i_1} - R^{i_1}\big)\dots\big(r^{i_n} - R^{i_n}\big) \big(r^i - R^i\big) \rho(\bm{r}),\nonumber \\
    \nu^{i_1 \dots i_n i} &= \frac{1}{c} \frac{n}{(n+1)!} \int d\bm{r}\, \big(r^{i_1} - R^{i_1}\big)\dots\big(r^{i_n} - R^{i_n}\big) \,\varepsilon^{iab} \big(r^a - R^a\big) j_P^b(\bm{r}).
\end{align}
We can write (with $k = \|\bm{k}\|$)
\begin{align}
    \Delta E_n^R = \frac{\hbar c}{\pi}\,\mathrm{p.v.}\sum_m E_{mn}^R \left(\frac{E_{mn}^R}{\hbar c}\right)^2 \sum_{a,b = 0}^{\infty} (+i)^a (-i)^b \left(\int_0^{\infty} dk \frac{k^{a+b}}{E_{mn}^R + \hbar ck }\right) \bigg[\Big(\mu_{nm}^{i_1\dots i_a i} \mathcal{I}_{(1)}^{i_1\dots i_a i j_1 \dots j_b j} \mu_{mn}^{j_1 \dots j_b j}\Big)\nonumber \\
    + \Big(\nu_{nm}^{i_1\dots i_a i} \mathcal{I}_{(1)}^{i_1\dots i_a i j_1 \dots j_b j} \nu_{mn}^{j_1 \dots j_b j}\Big) + 2\Im\Big(\mu_{nm}^{i_1\dots i_a i} \mathcal{I}_{(2)}^{i_1\dots i_a i j_1 \dots j_b j} \nu_{mn}^{j_1 \dots j_b j}\Big) \bigg].
\end{align}
\end{widetext}
Here we have defined the isotropic Cartesian tensors
\begin{align}
    \mathcal{I}^{i_1 \dots i_a i j_1\dots j_b j}_{(1)} &= \int \frac{d\Omega_{\bm{k}}}{4\pi}\, \Big(\delta^{ij} - \hat{k}^i \hat{k}^j\Big) \hat{k}^{i_1} \dots \hat{k}^{i_a} \hat{k}^{j_1} \dots \hat{k}^{j_b},\label{temp1000}\nonumber \\
    \mathcal{I}^{i_1 \dots i_a i j_1 \dots j_b j}_{(2)} &= \int \frac{d\Omega_{\bm{k}}}{4\pi}\, \Big(\varepsilon^{ijp} \hat{k}^p\Big) \hat{k}^{i_1} \dots \hat{k}^{i_a} \hat{k}^{j_1} \dots \hat{k}^{j_b}.
\end{align}
\\
The tensor $\mathcal{I}_{(1)}$ vanishes unless $a + b = 2n$ for some $n \in \mathbb{N}$, while the tensor $\mathcal{I}_{(2)}$ vanishes unless $a + b = 2n + 1$ for some $n \in \mathbb{N}$. These conditions preclude a number of electric, magnetic, and magnetoelectric terms, such as the product of the electric dipole and quadrupole moments, while allowing others like the product of the electric dipole and octopole moments. 

Unless $a=0$ and $b=0$ the integrals above diverge as $\|\bm{k}\|\to \infty$, even though the same expression written in terms of the current density was finite. This particular ultraviolet divergence 
is an artifact of our extrapolation of the Fourier transforms of the multipole expansions (\ref{fullmult}) to all values of $\|\bm{k}\|$. Indeed, for the straight-line paths considered above one can explicitly compute the Fourier transforms $\tilde{\bm{p}}(\bm{k})$ and $\tilde{\bm{m}}_{P}(\bm{k})$, after which one finds that they vanish in the limit $\|\bm{k}\|\to\infty$ \cite{Woolley4}. But the Fourier transforms of the multipole expansions (\ref{fullmult}) clearly diverge in the ultraviolet limit for all $n \in \mathbb{N}$, and so the resulting power series expansions in $\bm{k}$ must have a finite radius of convergence $\|\bm{k}\|\leq \Lambda'$ (not equal to the ultraviolet cutoff $\Lambda = (\hbar/mc)^{-1}$ used in previous sections). The exact value of $\Lambda'$ depends on the charge-current distribution, and it is therefore undesirable to regulate the Fourier integrals with a hard cutoff. We instead make these integrals well-behaved by introducing a ``heat kernel regulator" \cite{Schwartz, MolQED}, which in the continuum limit amounts to evaluating the integral 
    \begin{equation}
    \int_0^{\infty} dk \frac{k^{a+b}}{E_{mn}^R + \hbar ck}e^{- k / \Lambda'},
    \label{contint}
\end{equation}
where we take the limit $\Lambda' \to \infty$ to remove the regulator at the end of the calculation. This integral is evaluated in Appendix \ref{appendixf}. After removing the regulator, we find the general expression
\begin{widetext}
\begin{align}
    \Delta E_n^R = \frac{1}{\hbar \pi c} \sum_m E_{mn}^R \left(\frac{E_{mn}^R}{\hbar c}\right)^2 \ln(\frac{2mc^2}{\abs{E_{mn}^R}}) \sum_{a,b=0}^{\infty}(+i)^a (-i)^b (-1)^{a+b}\left(\frac{E_{mn}^R}{\hbar c}\right)^{a+b} \bigg[\Big(\mu_{nm}^{i_1\dots i_a i} \mathcal{I}_{(1)}^{i_1\dots i_a i j_1\dots j_b j} \mu_{mn}^{j_1\dots j_b j}\Big)\nonumber \\
    + \Big(\nu_{nm}^{i_1\dots i_a i} \mathcal{I}_{(1)}^{i_1\dots i_a i j_1\dots j_b j} \nu_{mn}^{j_1\dots j_b j}\Big) +2\Im\Big(\mu_{nm}^{i_1\dots i_a i}\mathcal{I}_{(2)}^{i_1\dots i_a i j_1\dots j_b j}\nu_{mn}^{j_1\dots j_b j}\Big)\bigg].\label{temp1000}
\end{align}
The first several contributions to $\Delta E_{n}^R$ are given by
\begin{align}
    \Delta E_n^R = \frac{2}{3 \hbar \pi c} \sum_m E_{mn}^R\left(\frac{E_{mn}^R}{\hbar c}\right)^2 \ln(\frac{2mc^2}{\abs{E_{mn}^R}}) \Bigg[\|\bm{\mu}_{nm}\|^2 + \|\bm{\nu}_{nm}\|^2 + \frac{1}{10}\left(\frac{E_{mn}^R}{\hbar c}\right)^2 \Big(3q_{nm}^{ab} q_{mn}^{ab} - q_{nm}^{aa} q_{mn}^{bb}\Big)\nonumber \\
    -\frac{1}{5}\left(\frac{E_{mn}^R}{\hbar c}\right)^2 \Big(\mu_{nm}^{a} o_{mn}^{abb} + o_{nm}^{abb} \mu_{mn}^a\Big) + \dots \Bigg],
\end{align}
\end{widetext}
where $\mu^{i}$, $q^{ij}$, and $o^{ijk}$ are the electric dipole, quadrupole, and octopole moment operators, $\nu^i$ is the magnetic dipole moment operator, and the ellipses denote the higher-order multipole contributions. To check the validity of this result, consider only the electric dipole term
\begin{equation}
    \Delta E_{n}^{R} = \frac{2}{3\hbar\pi c} \sum_m E_{mn}^R\left(\frac{E_{mn}^R}{\hbar c}\right)^2 \ln(\frac{2mc^2}{\abs{E_{mn}^R}}) \|\bm{\mu}_{nm}\|^2.
\end{equation}
\\
To recover Bethe's result in the electric dipole approximation, we first replace $\abs{E_{mn}^R}$ in the logarithm with an average value $\abs{\bar{E}_{n}^R}$, in which case
\begin{equation}
    \Delta E_{n}^{R} = \frac{2 \alpha}{3\pi} \frac{1}{m^2 c^2} \ln(\frac{2mc^2}{\abs{\bar{E}_n^R}}) \sum_m \big(E_{mn}^R\big)^3 \|\bm{\mu}_{nm}\|^2.
\end{equation}
In a first-quantized treatment this is precisely Bethe's result for the Lamb shift in the electric dipole approximation \cite{Bethe}. To compare with our previous result (\ref{eq5T11}), we use the identity \cite{MolQED}
\begin{equation}
    \big(E_{mn}^R\big)^2\|\bm{\mu}_{nm}\|^2 = - \|E_{mn}^R \bm{\mu}_{nm}\|^2 = \left(\frac{\hbar e}{m}\right)^2 \|\bm{\mathrm{P}}_{nm}\|^2,
\end{equation}
from which follows
\begin{equation}
    \Delta E_{n}^{R} = \frac{2 \alpha}{3\pi} \frac{1}{m^2 c^2} \ln(\frac{2mc^2}{\abs{\bar{E}_n^R}}) \sum_m E_{mn}^R \|\bm{\mathrm{P}}_{nm}\|^2.
\end{equation}
We recover the expression (\ref{eq5T11}) by evaluating the sum over states $\ket{\psi_m}$ using an expression analogous to (\ref{eq5T7}). 

\section{Conclusion}\label{section5}
We have introduced a field-theoretic reformulation of multipolar electrodynamics to model interactions between quantized electromagnetic fields and localized charge-current distributions. The electronic degrees of freedom are encoded in microscopic polarization and magnetization field operators --- defined through second-quantized scalar field operators --- with moments that are identified with the multipole moments of the charge-current distribution. These field operators couple to the quantized electromagnetic field in the multipolar Hamiltonian, 
obtained from the minimal coupling Hamiltonian through a unitary transformation, often referred to as the PZW transformation. Our reformulation generalizes existing work \cite{PowerTh1, PowerZienau} in that a field-theoretic model allows us to treat very general charge-current distributions --- provided they are sufficiently localized --- including those of large assemblies of atoms and molecules, and those in which the number of charge carriers may change over time. Restricting oneself to the electric dipole approximation is not required. 

Following a reformulation of multipolar electrodynamics from minimal coupling electrodynamics, 
our focus thereafter concerned the vacuum structure of the theory. 
Using Rayleigh-Schrödinger perturbation theory we computed the regularized shift $\Delta E_n^{\Lambda}$ of the electronic energy levels in the electromagnetic vacuum state, described by the regularized interaction term $H_{\text{int}}^{\Lambda}$. This energy shift 
depends explicitly on the ultraviolet cutoff $\Lambda$, an artifact of our choice of a ``hard cutoff" regulator. To remove this $\Lambda$-dependence, we renormalized the regularized Hamiltonian $H^{\Lambda}$ at leading-order in the fine structure constant by addition of appropriate counterterms. 
We could then use the renormalized Hamiltonian that followed to compute the \textit{finite} and observable correction $\Delta E_{n}^R$ to the (renormalized) electronic energy levels $E_n^R$. 

We obtain a closed form expression for the renormalized energy shift $\Delta E_n^R$ when written in terms of the matrix elements of the current density operator. For hydrogenic atoms, our expression reproduces existing results, including Bethe's calculation \cite{Bethe} of the Lamb shift in the electric dipole approximation, and extensions thereof \cite{Au,Grotch} to include spatial variations in the electromagnetic field. However, our expression (\ref{temp210}) is valid for more general assemblies of atoms and molecules (possibly with net charge), consisting of a sum of contributions centered on each ion in the assembly, each of which is weighted by the expectation value of the electronic charge density evaluated at that location. To compute these expectation values, one could expand the electron field operators in terms of an appropriate set of single-particle wavefunctions that transform in an irreducible representation of the symmetry group of the system \cite{Lax}. Because of how complicated this symmetry group may be, we would no longer expect the contributions to $\Delta E_n^R$ to be restricted to wavefunctions with full $\mathrm{SO}(3)$ symmetry like the $s$-type hydrogen orbitals; instead, we expect that these expectation values will depend in a complicated way on the electronic structure of the system. Numerically evaluating these expectation values and the expression for $\Delta E_n^R$ is an interesting direction for future work.

We then rewrote the renormalized energy shift in terms of the polarization and magnetization fields, permitting us to expand $\Delta E_n^R$ in a sum of contributions coming from products of specific electric and magnetic multipole moments of the molecule. We derived the general form of such expressions, and then identified the contributions coming from the first several mulitipole moments, namely the electric dipole, quadrupole, and magnetic dipole moments. Existing treatments of the vacuum energy shift have focused on hydrogenic atoms, and therefore include only the electric dipole term. Ours appears to be the 
first explicit expression for the renormalized energy shift that includes the full sum of multipole contributions --- they can now be investigated in detail.

This work lays the foundation for a broader research program involving microscopic polarization and magnetization fields in the quantum regime. 
Recent work indicates that a description of crystalline solids using multipolar quantum electrodynamics should be possible \cite{Swiecicki, Mahon1,Mahon2,Mahon3,Mahon4}; there the authors introduced a \textit{semi-classical} formalism based on microscopic polarization and magnetization fields that could be associated with individual lattice sites, together with itinerant contributions due to the presence of free charge and current. The extension of this into the fully quantum regime, by generalizing the work presented here to crystal lattices, should provide a microscopic underpinning for the study of quantum optical effects in crystals. 
\\
\begin{acknowledgements}
    This work was supported by the National Sciences and Engineering Research Council of Canada (NSERC).
\end{acknowledgements}

\appendix

\section{Transformations in Section \ref{section1}}
\subsection{Legendre transformation in minimal coupling}\label{appendixsum1}
We begin with the minimal coupling Lagrangian $L$ given in (\ref{eq2T1}). The canonical momentum density conjugate to the electron field $\psi(\bm{x},t)$ is
\begin{equation}
    \pi_{\psi}(\bm{x},t) = \frac{\delta L}{\delta \dot{\psi}(\bm{x},t)} = \frac{i \hbar}{2} \psi^{\dagger}(\bm{x},t),
    \label{eq2a4}
\end{equation}
while the canonical momentum density conjugate to the vector potential $\bm{a}(\bm{x},t)$ is
\begin{equation}
    \bm{\pi}_a(\bm{x},t) = \frac{\delta L}{\delta \dot{\bm{a}}(\bm{x},t)} = - \frac{1}{4\pi c} \bm{e}(\bm{x},t).\label{eq2atemp2}
\end{equation}
However, the canonical momentum density $\pi_{\phi}(\bm{x},t)$ conjugate to the scalar potential $\phi(\bm{x},t)$ vanishes. Thus the Lagrangian is degenerate, and the standard method for constructing the corresponding Hamiltonian theory involves the identification and classification of first- and second-class constraints using the Poisson bracket, elimination of the former by fixing a gauge, and enforcing the latter by replacing the Poisson bracket by a suitable generalization called the Dirac bracket \cite{Henneaux}. The solution in the case of electrodynamics is well-known, and so we 
only summarize the main results \cite{Woolley2, Woolley3}. There are two constraints: The vanishing of $\pi_{\phi}(\bm{x},t)$, and 
Gauss's law, $\bm{\nabla} \cdot \bm{e}(\bm{x},t) = 4\pi \rho^e(\bm{x},t)$. After fixing the transverse gauge
\begin{equation}
    \bm{\nabla} \cdot \bm{a}(\bm{x},t) = 0,\label{eq2a16}
\end{equation}
the scalar potential is obtained from the electronic charge density through Poisson's equation
\begin{equation}
    \nabla^2 \phi(\bm{x},t) = 4\pi \rho^e(\bm{x},t)\label{eq2T2},
\end{equation}
and the canonical variable $\pi_{\phi}(\bm{x},t)$ can be discarded. The \textit{gauge-fixed} Maxwell Lagrangian $L_B'$ is
\begin{equation}
    L_B' = \frac{1}{8\pi} \int d\bm{x} \left(\frac{1}{c^2}\|\dot{\bm{a}}(\bm{x},t)\|^2 - \|\bm{\nabla}\times\bm{a}(\bm{x},t)\|^2\right),
\end{equation}
while the gauge-fixed interaction term is
\begin{equation}
    L_{\text{int}}' = L_{\text{int}} + \frac{1}{2} \int d\bm{x}\, \rho^e(\bm{x},t)\phi(\bm{x},t).
\end{equation}
Using Poisson's equation, the contribution to the Maxwell Lagrangian $L_B$ coming from the longitudinal electric field (absent in the gauge-fixed $L_B'$) is rewritten in terms of the electronic charge density and scalar potential; this is the origin of the factor $1/2$ in the second term above. After gauge-fixing, Gauss's law becomes a second-class constraint and is enforced upon quantization by replacing Dirac brackets of the remaining canonical variables with (anti)commutators, the result being the equal-time commutation relations (\ref{eq2a17}) for the electromagnetic field operators and the equal-time anticommutation relations (\ref{eq2a10}) for the electron field operators. Then the mimimal coupling Hamiltonian (\ref{eq2a19}) follows. 

\subsection{The PZW transformation}\label{appendixsum2}
We begin with the minimal coupling Hamiltonian (\ref{eq2a19}), which can be written as a functional of the canonical variables
\begin{equation}
    H = \mathscr{F}\big[\bm{a},\bm{\pi}_{a}, \psi,\pi_{\psi}\big],
    \label{eq2b8}
\end{equation}
where $\bm{\pi}_a(\bm{x})$ and $\pi_{\psi}(\bm{x})$ are the canonical momentum densities defined in (\ref{eq2atemp2}) and (\ref{eq2a4}), respectively. Under the unitary transformation $\mathrm{U} \equiv \exp(iS/\hbar)$, the new canonical variables (indicated by a breve) are
\begin{align}
    \Breve{\bm{a}}(\bm{x}) &= \mathrm{U} \bm{a}(\bm{x}) \mathrm{U}^{\dagger},\nonumber \\
    \Breve{\bm{\pi}}_a(\bm{x}) &= \mathrm{U} \bm{\pi}_a(\bm{x}) \mathrm{U}^{\dagger},\nonumber \\
    \Breve{\psi}(\bm{x}) &= \mathrm{U} \psi(\bm{x}) \mathrm{U}^{\dagger},\nonumber \\
    \Breve{\pi}_{\psi}(\bm{x}) &= \mathrm{U} \pi_{\psi}(\bm{x}) \mathrm{U}^{\dagger}.\label{eq2b12}
\end{align}
The strategy then \cite{Healy} is to write the Hamiltonian as a functional of the \textit{transformed} fields,
\begin{equation}
    H = \mathscr{G}\big[\Breve{\bm{a}},\Breve{\bm{\pi}}_{a},\Breve{\psi},\Breve{\pi}_{\psi}\big]\label{eq2b13}.
\end{equation}
Equating (\ref{eq2b8}) and $(\ref{eq2b13})$ and using the general form of the transformations (\ref{eq2b12}), the new functional $\mathscr{G}$ is obtained from the old one $\mathscr{F}$ through
\begin{equation}
    \mathscr{G}\big[\bm{a},\bm{\pi}_{a}, \psi,\pi_{\psi}\big] = \mathrm{U}^{\dagger} \mathscr{F}\big[\bm{a},\bm{\pi}_{a}, \psi,\pi_{\psi}\big] \mathrm{U},
\end{equation}
where on both sides we use the \textit{non-transformed fields}. Once the new functional $\mathscr{G}$ is identified, the multipolar Hamiltonian is obtained by substituting the transformed field operators into this new functional. The explicit transformations of the field operators are
\begin{align}
    \Breve{\bm{a}}(\bm{x}) &= \bm{a}(\bm{x}),\nonumber \\
    \Breve{\bm{\pi}}_a(\bm{x}) &= \bm{\pi}_a(\bm{x}) - \frac{1}{c} \bm{p}_T(\bm{x}),\nonumber \\
    \Breve{\psi}(\bm{x}) &= e^{- i \Phi(\bm{x},\bm{R})} \psi(\bm{x}),\nonumber \\
    \Breve{\pi}_{\psi}(\bm{x}) &= e^{+ i \Phi(\bm{x},\bm{R})} \pi_{\psi}(\bm{x}),\label{eq2b18}
\end{align}
where
\begin{equation}
    \Phi(\bm{x},\bm{R}) = \frac{e}{\hbar c} \int d\bm{w}\, s^i(\bm{w};\bm{x},\bm{R}) a^i(\bm{w}).\label{eq2b19}
\end{equation}
In 
semiclassical theory the quantity $\Phi(\bm{x},\bm{R})$ is a generalized \textit{Peierls phase} \cite{Mahon1}; in the fully quantum theory, the exponential of this quantity in (\ref{eq2b18}) is the \textit{Wilson line operator} of Abelian gauge theory \cite{Woolley4}. The original gauge freedom involving the scalar and vector potentials has been replaced \cite{Woolley4} by a freedom in choosing the relators ((\ref{eq2b7}) and (\ref{eq2b31})); here this involves a choice of the paths $C(\bm{y},\bm{R})$, although other classes of relators are possible \cite{Sipeold}.

With the canonical momentum density conjugate to the vector potential given in (\ref{eq2atemp2}), the transformed transverse electric field is $\Breve{\bm{e}}_T(\bm{x}) = \bm{e}_T(\bm{x}) + 4\pi \bm{p}_T(\bm{x})$, which we identify with the transverse part of the displacement field $\bm{d}_T(\bm{x})$, while the magnetic field is unchanged, $\Breve{\bm{b}}(\bm{x}) = \bm{b}(\bm{x})$. Then, after dropping the breve accent on the transformed electron field operators, the multipolar Hamiltonian (\ref{temp1}) follows.

\begin{widetext}
\section{Electrostatic interactions in the multipolar Hamiltonian}\label{appendixA}
We demonstrate the equivalence between the regularized form of (\ref{temp3}) and (\ref{temp14}). To begin, we use \cite{MolQED}
\begin{align}
    \delta^{ij}_{L}(\bm{x}-\bm{x}') &= \int \frac{d\bm{k}}{(2\pi)^3} \hat{k}^i \hat{k}^j e^{i\bm{k}\cdot(\bm{x}-\bm{x}')} = - \frac{1}{2\pi^2} \int_0^{\infty} dk\, \partial^i \partial^j F(\bm{x},\bm{x}';k),\label{long}
\end{align}
where $F(\bm{x},\bm{x}';k)$ is given by (\ref{temp8}). After regularizing the Fourier integrals in the representation (\ref{long}) of the longitudinal delta functions, the second term on the first line of (\ref{temp3}) is
\begin{align}
    2\pi\int d\bm{x}\, \|\bm{p}_{L}(\bm{x})\|^2 &= \frac{1}{\pi} \int_0^{\Lambda} dk \iint d\bm{x}d\bm{x}'\,\big(\rho(\bm{x}')-\rho_F(\bm{x}')\big)F(\bm{x},\bm{x}';k)\big(\rho(\bm{x})-\rho_F(\bm{x})\big),\nonumber \\
    &= \frac{1}{\pi} \int_0^{\Lambda} dk \iint d\bm{x}d\bm{x}'\,\rho(\bm{x}')F(\bm{x},\bm{x}';k)\rho(\bm{x}) - \frac{2Q}{\pi}\int_0^{\Lambda} dk \int d\bm{x}\, \rho(\bm{x}) F(\bm{x},\bm{R};k) + \frac{Q^2}{\pi} \int_0^{\Lambda} dk.
\end{align}
Meanwhile, using Gauss's law $\bm{\nabla}\cdot\bm{d}_{L}(\bm{x}) = 4\pi\rho_{F}(\bm{x})$, a straightforward calculation leads to
\begin{equation}
    - \int d\bm{x}\, \bm{d}_{L}(\bm{x})\cdot\left(\bm{p}(\bm{x}) - \frac{1}{8\pi} \bm{d}_{L}(\bm{x})\right) = - \frac{Q^2}{\pi} \int_0^{\Lambda} dk + \frac{2Q}{\pi} \int_0^{\Lambda} dk \int d\bm{x} \rho(\bm{x}) F(\bm{x}, \bm{R};k),
\end{equation}
and therefore
\begin{align}
    H_F = \frac{\hbar^2}{2m} \int d\bm{x}\, \bm{\nabla}\psi^{\dagger}(\bm{x})\cdot\bm{\nabla}\psi(\bm{x}) + \frac{1}{\pi} \int_0^{\Lambda} dk \iint d\bm{x}d\bm{x}' \rho(\bm{x}')F(\bm{x},\bm{x}';k)\rho(\bm{x}).
\end{align}
Next, using $\rho(\bm{x}) = \rho^e(\bm{x}) + \rho^{\text{ion}}(\bm{x})$, 
\begin{align}
    \frac{1}{\pi} \int_0^{\Lambda} dk \iint d\bm{x}d\bm{x}' \rho(\bm{x}')F(\bm{x},\bm{x}';k)\rho(\bm{x}) = \frac{1}{\pi} \int_0^{\Lambda} dk \iint d\bm{x}d\bm{x}' \rho^e(\bm{x}')F(\bm{x},\bm{x}';k)\rho^e(\bm{x})\nonumber \\
    + \frac{2}{\pi} \sum_N q_N \int_0^{\Lambda} dk \int d\bm{x}\, F(\bm{x},\bm{R}+\bm{d}_N;k) \rho^e(\bm{x}) + \sum_{NM} \frac{q_N q_M}{\pi} \int_0^{\Lambda} dk\, F(\bm{d}_N, \bm{d}_{M};k),
\end{align}
and so, isolating the divergences in $E_{0,F}^{\Lambda}$, we have
\begin{align}
    \frac{1}{\pi} \int_0^{\Lambda} dk \iint d\bm{x}d\bm{x}' \rho(\bm{x}')F(\bm{x},\bm{x}';k)\rho(\bm{x}) = \frac{1}{\pi} \int_0^{\Lambda} dk \iint d\bm{x}d\bm{x}' \psi^{\dagger}(\bm{x})\psi^{\dagger}(\bm{x}')F(\bm{x},\bm{x}';k)\psi(\bm{x}')\psi(\bm{x})\nonumber \\
    + \frac{2}{\pi} \sum_N q_N \int_0^{\Lambda} dk \int d\bm{x}\, F(\bm{x},\bm{R}+\bm{d}_N;k) \rho^e(\bm{x}) + \sum_{N\neq M} \frac{q_N q_M}{\pi} \int_0^{\Lambda} dk\, F(\bm{d}_N, \bm{d}_{M};k) + E_{0,F}^{\Lambda},
\end{align}
where in the integral in the second term on the right side of the above equality it should be understood that a small sphere centered at $\bm{x} = \bm{R} + \bm{d}_N$ is excluded; the $\Lambda$-dependent term is given by
\begin{equation}
    E_{0,F}^{\Lambda} = \frac{1}{\pi} \left[e^2 N_e + 2 e N_e \sum_{N} q_N +\sum_N q_N^2\right] \Lambda,
\end{equation}
and the second term on the right-hand-side of this expression comes from the integration over that small sphere. Since we have isolated the $\Lambda$-dependent terms in $E_{0,F}^{\Lambda}$, we can take the limit $\Lambda \to \infty$ in the remaining terms and use
\begin{equation}
    \int_0^{\infty} dk\, F(\bm{x},\bm{x}';k) = \frac{\pi}{2} \frac{1}{\|\bm{x}-\bm{x}'\|},
\end{equation}
which is valid for $\bm{x}\neq\bm{x}'$. Dropping the ion-ion interaction as in Section \ref{section1}, we obtain the desired result (\ref{temp14}).
\end{widetext}

\section{First-order diamagnetic contribution}\label{appendixb}
Here we simplify the diamagnetic contribution to $\Delta E_{n(1)}^{\Lambda}$. The diamagnetic term is
\\
\begin{equation}
    H_{D} \equiv -\frac{1}{2} \int d\bm{x}\, \bm{m}_D(\bm{x})\cdot\bm{b}(\bm{x}).
\end{equation}
\\
As noted in the main text, we can rewrite $H_D$ explicitly as (\ref{temporary1}); using (\ref{eq2a14}) we write this as
\\
\begin{equation}
    H_D = \frac{1}{2} \iint d\bm{x}d\bm{x}'\, a^a(\bm{x}) \Breve{O}^{ab}(\bm{x},\bm{x}') a^b(\bm{x}'),\label{b1}
\end{equation}
\\
where
\begin{equation}
    \Breve{O}^{ab}(\bm{x},\bm{x}') = \varepsilon^{api}\varepsilon^{b\ell j} \frac{\partial}{\partial x^p} \frac{\partial}{\partial x'^{\ell}} O^{ij}(\bm{x},\bm{x}').
\end{equation}
and where $O^{ij}(\bm{x},\bm{x}')$, the second-quantized diamagnetization field, is defined in (\ref{temporary2}). With \cite{Mahon1}
\\
\begin{equation}
    \varepsilon^{abc} \frac{\partial}{\partial x^b} \alpha^{cd}(\bm{x};\bm{y},\bm{R}) = - \frac{\partial}{\partial y^a} s^d(\bm{x};\bm{y},\bm{R}) + \delta^{ad} \delta(\bm{x}-\bm{y})
\end{equation}
\\
we have
\begin{widetext}
\begin{align}
    \Breve{O}^{ab}(\bm{x},\bm{x}') = \frac{e}{2mc^2} \int d\bm{y} \bigg(\delta^{ab} \delta(\bm{x}-\bm{y})\delta(\bm{x}'-\bm{y}) - \delta^{ia} \delta(\bm{x}-\bm{y}) \frac{\partial}{\partial y^b} s^i(\bm{x}';\bm{y},\bm{R}) - \delta^{ib}\delta(\bm{x}'-\bm{y}) \frac{\partial}{\partial y^a} s^i(\bm{x};\bm{y},\bm{R})\nonumber \\
    + \left[\frac{\partial}{\partial y^a} s^i(\bm{x};\bm{y},\bm{R})\right]\left[\frac{\partial}{\partial y^b} s^i(\bm{x}';\bm{y},\bm{R})\right]\bigg) \rho^e(\bm{y}).
\end{align}
Substituting this expression back into (\ref{b1}), the diamagnetic term is
\begin{equation}
    H_{D} = \frac{e}{2mc^2} \int d\bm{x}\, \rho^e(\bm{x}) \|\bm{a}(\bm{x})\|^2 - \frac{\hbar}{mc} \int d\bm{x}\, \rho^e(\bm{x}) \bm{a}(\bm{x})\cdot\bm{\nabla}\Phi(\bm{x},\bm{R}) + \frac{\hbar^2}{2me} \int d\bm{x}\, \rho^e(\bm{x}) \|\bm{\nabla}\Phi(\bm{x},\bm{R})\|^2.\label{b2}
\end{equation}
\end{widetext}
Define
\begin{equation}
    \Delta E_{D}^{\Lambda} \equiv \bra{\Psi_n} H_{D}^{\Lambda}\ket{\Psi_n},
\end{equation}
with $\ket{\Psi_n}$ given by (\ref{eq3c1}). The regularized mode expansion for the vector potential $\bm{a}(\bm{x})$ that leads to the mode expansion (\ref{eq3a13}) for $\bm{b}(\bm{x})$ is
\begin{equation}
    \bm{a}(\bm{x}) = \sum\limits_{I}\sum_{\|\bm{k}\|\leq \Lambda}\left(\frac{2\pi \hbar c}{V \|\bm{k}\|}\right)^{1/2} \bm{e}_{I\bm{k}} a_{I\bm{k}} e^{i\bm{k}\cdot\bm{x}} + \text{h.c.}.\label{eqTempT}
\end{equation}
A straightforward calculation of the vacuum expectation value of $\|\bm{a}(\bm{x})\|^2$ leads to
\begin{equation}
    \frac{e}{2mc^2} \int d\bm{x}\, \rho_{nn}^e(\bm{x})\expval{\|\bm{a}(\bm{x})\|^2}_{\text{vac}} = \frac{\alpha}{2\pi}\left(\frac{\hbar^2 N_e}{m}\right) \Lambda^2.
\end{equation}
Focusing on latter two terms of (\ref{b2}), the expectation value of the first is
\begin{equation}
    - \frac{\hbar}{mc} \int d\bm{x}\, \rho_{nn}^e(\bm{x}) \expval{\bm{a}(\bm{x})\cdot\bm{\nabla}\Phi(\bm{x},\bm{R})}_{\text{vac}},\label{b3}
\end{equation}
while the expectation value of the second is 
\begin{equation}
    +\frac{\hbar^2}{2me} \int d\bm{x}\, \rho_{nn}^e(\bm{x})\expval{\|\bm{\nabla}\Phi(\bm{x},\bm{R})\|^2}_{\text{vac}}.\label{b4}
\end{equation}
With the regularized mode expansion for $\bm{a}(\bm{x})$, in the continuum limit we have
\begin{equation}
    \expval{a^a(\bm{x})a^b(\bm{x}')}_{\text{vac}} = \frac{\hbar c}{4\pi^2} \int_{\|\bm{k}\|\leq \Lambda} d\bm{k}\, \|\bm{k}\|^{-1} \delta_T^{ab}(\bm{k}) e^{i\bm{k}\cdot(\bm{x}'-\bm{x})}.
\end{equation}
Thus, the vacuum expectation value in (\ref{b3}) is
\begin{widetext}
\begin{align}
- \frac{\hbar e}{4\pi^2 mc} \int_{\|\bm{k}\|\leq \Lambda} d\bm{k}\|\bm{k}\|^{-1} \delta_T^{ij}(\bm{k}) \iint d\bm{x}d\bm{x}'\, e^{i\bm{k}\cdot(\bm{x}'-\bm{x})} \rho_{nn}^e(\bm{x}) \left[\frac{\partial}{\partial x^i} s^j(\bm{x}';\bm{x},\bm{R})\right],
\end{align}
while the vacuum expectation value in (\ref{b4}) is
\begin{align}
    \frac{\hbar e}{8\pi^2 m c} \int_{\|\bm{k}\|\leq \Lambda} d\bm{k}\|\bm{k}\|^{-1} \delta_T^{ij}(\bm{k}) \iint d\bm{x}d\bm{x}'\, e^{i\bm{k}\cdot(\bm{x}'-\bm{x})} \int d\bm{y}\, \rho_{nn}^e(\bm{y}) \left[\frac{\partial}{\partial y^k} s^i(\bm{x};\bm{y},\bm{R})\right] \left[\frac{\partial}{\partial y^k} s^j(\bm{x}';\bm{y},\bm{R})\right],
\end{align}
so that the expectation values of the latter two terms of (\ref{b2}) are
\begin{align}
    - \frac{\hbar}{mc} \int d\bm{x}\, \rho_{nn}^e(\bm{x}) \expval{\bm{a}(\bm{x})\cdot\bm{\nabla}\Phi(\bm{x},\bm{R})}_{\text{vac}} + \frac{\hbar^2}{2me} \int d\bm{x}\, \rho_{nn}^e(\bm{x}) \expval{\|\bm{\nabla}\Phi(\bm{x},\bm{R})\|^2}_{\text{vac}} = - \frac{\hbar e}{4\pi^2 mc} \int_{\|\bm{k}\|\leq \Lambda} d\bm{k}\|\bm{k}\|^{-1} \delta_T^{ij}(\bm{k})\nonumber \\
    \times \iint d\bm{x}d\bm{x}'\, e^{i\bm{k}\cdot(\bm{x}'-\bm{x})} \bigg(\rho_{nn}^e(\bm{x})\left[\frac{\partial}{\partial x^i} s^j(\bm{x}';\bm{x},\bm{R})\right] - \frac{1}{2} \int d\bm{y}\, \rho_{nn}^e(\bm{y})\left[\frac{\partial}{\partial y^k} s^i(\bm{x};\bm{y},\bm{R})\right]\left[\frac{\partial}{\partial y^k} s^j(\bm{x}';\bm{y},\bm{R})\right]\bigg).\label{cf1}
\end{align}
In total, the diamagnetic contribution to the first-order correction is then (\ref{eq3c15}).
\end{widetext}

\section{Second-order correction}
\subsection{A useful identity}\label{appendixc1}
We first prove a useful identity concerning matrix elements of field operators. Let $\mathcal{O}(\bm{x})$ denote a field operator in the Schrödinger picture which acts trivially on the Hilbert space $\mathcal{H}_B$ of the Bose sector. Separating the free and interaction terms of the multipolar Hamiltonian as in (\ref{eq3a26}), we define an interaction picture representation by taking the time-dependence of the field operator in the interaction picture to be
\begin{equation}
    \mathcal{O}_I(\bm{x},t) = e^{i H_0 t / \hbar} \mathcal{O}(\bm{x}) e^{-i H_0 t / \hbar},
    \label{c1}
\end{equation}
where
\begin{equation}
    H_0 = H_B \otimes \mathbb{I}_F + \mathbb{I}_B \otimes H_F
\end{equation}
with $H_B$ and $H_F$ given by (\ref{temp2}) and (\ref{temp3}), while an interaction picture state $\ket{\Psi_{n}}_{I}$ is related to the corresponding Schrödinger picture state $\ket{\Psi_n}_S$ through
\begin{equation}
    \ket{\Psi_n}_I = e^{i H_{0} t / \hbar}\ket{\Psi_n}_S.
    \label{c2}
\end{equation}
The expression (\ref{c1}) is equivalent to
\begin{equation}
    \frac{\partial\mathcal{O}_I(\bm{x},t)}{\partial t} = \frac{1}{i \hbar} \big[\mathcal{O}_I(\bm{x},t), H_0\big]_-.
    \label{c3}
\end{equation}
Since $\mathcal{O}(\bm{x})$ acts non-trivially only on $\mathcal{H}_F$ by assumption, it follows that the commutator above also acts non-trivially only on $\mathcal{H}_F$, so that
\begin{align}
    \frac{\partial\mathcal{O}_{I,nm}(\bm{x},t)}{\partial t} &= \frac{1}{i\hbar} \bra{\psi_n}\big[\mathcal{O}_I(\bm{x},t), H_0\big]_-\ket{\psi_m}\nonumber \\
    &= (i\hbar)^{-1} E_{mn} \mathcal{O}_{I,nm}(\bm{x},t),
    \label{c4}
\end{align}
where the subscript ``$nm$" denotes the matrix element between the eigenstates $\ket{\psi_{n,m}}$ of $H_F$, and $E_{mn} = E_m - E_n$ with $E_{n,m}$ being the eigenvalues of $H_F$ corresponding to the eigenstates $\ket{\psi_{n,m}}$. Taking $\mathcal{O}(\bm{x})$ to be the paramagnetic current density operator $\bm{j}_P(\bm{x})$ we have
\begin{align}
    \bm{j}_{P,nm}(\bm{x}) = \bra{\psi_n}\bm{j}_P(\bm{x})\ket{\psi_m}_S = \bra{\psi_n} \bm{j}_{P,I}(\bm{x},t)\ket{\psi_m}_I. 
    \label{c5}
\end{align}
In the interaction picture we can split the Heisenberg-picture equation (\ref{temp6}) into a pair of equations
\begin{align}
    \bm{j}_{P,I}(\bm{x},t) &= \frac{\partial \bm{p}_I(\bm{x},t)}{\partial t} + c \bm{\nabla}\times\bm{m}_{P,I}(\bm{x},t),\label{c6} \\
    \bm{j}_{D,I}(\bm{x},t) &= c\bm{\nabla}\times\bm{m}_{D,I}(\bm{x},t).\label{c7}
\end{align}
Through the expressions (\ref{c4}-\ref{c6}) we thereby obtain the Schrödinger-picture identity
\begin{equation}
    \bm{j}_{P,nm}(\bm{x}) = (i\hbar)^{-1} E_{mn} \bm{p}_{nm}(\bm{x}) + c\bm{\nabla}\times\bm{m}_{P,nm}(\bm{x}),\label{c8}
\end{equation}
which is used in the main text and below.

\begin{widetext}
\subsection{Simplifications}\label{appendixc}
Begin with the full expression for the second-order correction (\ref{eq3d5}). We can simplify the second line using \cite{MolQED}
\begin{equation}
    \frac{1}{4\pi} \int d\Omega_{\bm{k}}\,\varepsilon^{ipj} \frac{k^p}{\|\bm{k}\|} e^{i \bm{k}\cdot(\bm{x}'-\bm{x})} = \frac{i}{\|\bm{k}\|} \varepsilon^{ipj}\partial'^p F(\bm{x},\bm{x}';k),\label{eq3d6}
\end{equation}
where $F(\bm{x},\bm{x}';k)$ is given by (\ref{temp8}). Together with the identity (\ref{eq3c9}), we have
    \begin{align}
         \Delta E_{n(2)}^{\Lambda} = &- \frac{1}{\pi}\,\mathrm{p.v.} \sum_{m} \int_0^{\Lambda} dk \frac{\hbar c k}{E_{mn} + \hbar c k}\, \iint d\bm{x}d\bm{x}'\, \tau^{ij}\big(k\|\bm{x}'-\bm{x}\|\big) \Big(p_{nm}^i(\bm{x}') p_{mn}^j(\bm{x}) + m_{P,nm}^i(\bm{x}') m_{P,mn}^j(\bm{x})\Big)\nonumber
    \\
     &- \frac{i}{\pi}\,\mathrm{p.v.} \sum_{m} \int_0^{\Lambda} dk \frac{\hbar c k^2}{E_{mn} + \hbar c k}\, \iint d\bm{x}d\bm{x}'\, \sigma^{ij}\big(k\|\bm{x}'-\bm{x}\|\big) \Big(p_{nm}^i(\bm{x}') m_{P,mn}^j(\bm{x}) + m_{P,nm}^i(\bm{x}') p_{mn}^j(\bm{x})\Big).\label{eq3d7}
    \end{align}
To simplify $\Delta E_{n(2)}^{\Lambda}$, we break up its summands and process them in turn. We denote by $\Delta E_{\mathrm{pp}}^{\Lambda}$ the first term involving the product of polarization fields, by $\Delta E_{\mathrm{mm}}^{\Lambda}$ the second term involving the product of the magnetization fields, and by $\Delta E_{\mathrm{pm}}^{\Lambda}$ the second line above. Using twice the algebraic identity 
\begin{equation}
    \frac{1}{E_{mn} + \hbar c k} = \frac{1}{\hbar ck} - \frac{1}{\hbar ck} \frac{E_{mn}}{E_{mn} + \hbar ck},\label{temp10}
\end{equation}
we have
    \begin{align}
        \Delta E_{\mathrm{pp}}^{\Lambda} \equiv &- \frac{1}{\pi}\,\mathrm{p.v.} \sum_{m} \int_0^{\Lambda} dk \frac{\hbar c k}{E_{mn} + \hbar c k}\, \iint d\bm{x}d\bm{x}'\, \tau^{ij}\big(k\|\bm{x}'-\bm{x}\|\big) p_{nm}^i(\bm{x}') p_{mn}^j(\bm{x})\nonumber \\
        = &-\frac{1}{\pi} \sum_{m} \int_0^{\Lambda} dk\, \iint d\bm{x} d\bm{x}'\, \tau^{ij}\big(k\|\bm{x}'-\bm{x}\|\big) p_{nm}^i(\bm{x}') p_{mn}^j(\bm{x})\nonumber \\
        &+\frac{1}{\hbar \pi c} \sum_{m} \int_0^{\Lambda} dkk^{-1}\,E_{mn} \iint d\bm{x} d\bm{x}'\, \tau^{ij}\big(k\|\bm{x}'-\bm{x}\|\big) p_{nm}^i(\bm{x}') p_{mn}^j(\bm{x})\nonumber \\
        &-\frac{1}{\hbar \pi c}\,\mathrm{p.v.} \sum_{m} \int_0^{\Lambda} dk k^{-1}\, \frac{\big(E_{mn}\big)^2}{E_{mn} + \hbar c k} \iint d\bm{x}d\bm{x}'\, \tau^{ij}\big(k\|\bm{x}'-\bm{x}\|\big) p_{nm}^i(\bm{x}') p_{mn}^j(\bm{x}).\label{eq3d15}
    \end{align}
Using our identity (\ref{c8}) this can be written as
    \begin{align}
        \Delta E_{\mathrm{pp}}^{\Lambda} = &-\frac{1}{\pi} \sum_{m} \int_0^{\Lambda} dk\, \iint d\bm{x}d\bm{x}'\, \tau^{ij}\big(k\|\bm{x}'-\bm{x}\|\big) p_{nm}^i(\bm{x}') p_{mn}^j(\bm{x})\nonumber \\
        &+\frac{1}{\hbar \pi c} \sum_{m} \int_0^{\Lambda} dk k^{-1} \iint d\bm{x}d\bm{x}'\, \tau^{ij}\big(k\|\bm{x}'-\bm{x}\|\big) E_{mn} p_{nm}^i(\bm{x}') p_{mn}^j(\bm{x})\nonumber \\
        &-\frac{1}{\pi}\mathrm{p.v.} \sum_{m} \int_0^{\Lambda} dk\, \frac{\hbar c k}{E_{mn} + \hbar c k} \iint d\bm{x} d\bm{x}'\, \tau^{ij}\big(k\|\bm{x}'-\bm{x}\|\big) m_{P,nm}^i(\bm{x}') m_{P,mn}^j(\bm{x})\nonumber \\
        &-\frac{\hbar}{\pi c}\,\mathrm{p.v.} \sum_{m} \int_0^{\Lambda} dk\, \frac{k^{-1}}{E_{mn} + \hbar c k} \iint d\bm{x} d\bm{x}'\, \tau^{ij}\big(k\|\bm{x}'-\bm{x}\|\big) j_{P,nm}^i(\bm{x}') j_{P,mn}^j(\bm{x})\nonumber \\
        &+\frac{\hbar}{\pi}\,\mathrm{p.v.} \sum_{m} \int_0^{\Lambda} dk\, \frac{k}{E_{mn} + \hbar c k} \iint d\bm{x} d\bm{x}'\, \sigma^{ij}\big(k\|\bm{x}'-\bm{x}\|\big)\Big(j_{P,nm}^i(\bm{x}') m_{P,mn}^j(\bm{x}) + m_{P,nm}^i(\bm{x}') j_{P,mn}^j(\bm{x})\Big).\label{eq3d15}
    \end{align}
We can combine the third line above with the contribution
\begin{align}
    \Delta E_{\mathrm{mm}}^{\Lambda} = &- \frac{1}{\pi}\,\mathrm{p.v.} \sum_{m} \int_0^{\Lambda} dk \frac{\hbar c k}{E_{mn} + \hbar c k}\, \iint d\bm{x}d\bm{x}'\, \tau^{ij}\big(k\|\bm{x}'-\bm{x}\|\big) m_{nm}^i(\bm{x}') m_{mn}^j(\bm{x}).\label{mm}
\end{align}
Consider the last line of (\ref{eq3d7}). Using the algebraic identity (\ref{temp10}) along with the identity (\ref{c8}), that term can be written
    \begin{align}
        \Delta E_{\mathrm{pm}}^{\Lambda} &\equiv - \frac{i}{\pi}\,\mathrm{p.v.} \sum_{m} \int_0^{\Lambda} dk \frac{\hbar c k^2}{E_{mn} + \hbar c k}\, \iint d\bm{x}d\bm{x}'\, \sigma^{ij}\big(k\|\bm{x}'-\bm{x}\|\big) \Big(p_{nm}^i(\bm{x}') m_{P,mn}^j(\bm{x}) + m_{P,nm}^i(\bm{x}') p_{mn}^j(\bm{x})\Big)\nonumber \\
        &= - \frac{i}{\pi} \int_0^{\Lambda} dkk \iint d\bm{x}d\bm{x}'\, \sigma^{ij}\big(k\|\bm{x}'-\bm{x}\|\big) \expval{p^i(\bm{x}')m_P^j(\bm{x}) - m_P^i(\bm{x}') p^j(\bm{x})}_n\nonumber \\
        &+\frac{2}{\pi}\,\mathrm{p.v.} \sum_{m} \int_0^{\Lambda} dk\, \frac{\hbar c k}{E_{mn} + \hbar c k} \iint d\bm{x} d\bm{x}'\, \tau^{ij}\big(k\|\bm{x}'-\bm{x}\|\big) m_{P,nm}^i(\bm{x}') m_{P,mn}^j(\bm{x})\nonumber \\
        &-\frac{\hbar}{\pi}\,\mathrm{p.v.} \sum_{m} \int_0^{\Lambda} dk\, \frac{k}{E_{mn} + \hbar ck} \iint d\bm{x}d\bm{x}'\, \sigma^{ij}\big(k\|\bm{x}'-\bm{x}\|\big) \Big(j_{P,nm}^i(\bm{x}')m_{P,mn}^j(\bm{x}) + m_{P,nm}^j(\bm{x}') j_{P,mn}^i(\bm{x})\Big).
    \end{align}
Then, through a trivial relabelling
    \begin{align}
        \Delta E_{\mathrm{pm}}^{\Lambda} &= - \frac{i}{\pi} \int_0^{\Lambda} dkk \iint d\bm{x}d\bm{x}'\, \sigma^{jk}\big(k\|\bm{x}'-\bm{x}\|\big) \expval{\big[m_P^k(\bm{x}'),p^j(\bm{x})\big]_-}_n\nonumber \\
        &+\frac{2}{\pi}\,\mathrm{p.v.} \sum_{m} \int_0^{\Lambda} dk\, \frac{\hbar c k}{E_{mn} + \hbar c k} \iint d\bm{x} d\bm{x}'\, \tau^{ij}\big(k\|\bm{x}'-\bm{x}\|\big) m_{P,nm}^i(\bm{x}') m_{P,mn}^j(\bm{x})\nonumber \\
        &-\frac{\hbar}{\pi}\,\mathrm{p.v.} \sum_{m} \int_0^{\Lambda} dk\, \frac{k}{E_{mn} + \hbar ck} \iint d\bm{x}d\bm{x}'\, \sigma^{ij}\big(k\|\bm{x}'-\bm{x}\|\big) \Big(j_{P,nm}^i(\bm{x}')m_{P,mn}^j(\bm{x}) + m_{P,nm}^j(\bm{x}') j_{P,mn}^i(\bm{x})\Big).\label{pm}
    \end{align}
With the expressions (\ref{eq3d15}) for $\Delta E_{\mathrm{pp}}^{\Lambda}$, (\ref{mm}) for $\Delta E_{\mathrm{mm}}^{\Lambda}$, and (\ref{pm}) for $\Delta E_{\mathrm{pm}}^{\Lambda}$, the total second-order correction is
    \begin{align}
        \Delta E_{n(2)}^{\Lambda} = &-\frac{1}{\pi} \sum_{m} \int_0^{\Lambda} dk\, \iint d\bm{x}d\bm{x}'\, \tau^{ij}\big(k\|\bm{x}'-\bm{x}\|\big) p_{nm}^i(\bm{x}') p_{mn}^j(\bm{x})\nonumber \\
        &+\frac{1}{\hbar \pi c} \sum_{m} \int_0^{\Lambda} dk k^{-1} \iint d\bm{x}d\bm{x}'\, \tau^{ij}\big(k\|\bm{x}'-\bm{x}\|\big) E_{mn} p_{nm}^i(\bm{x}') p_{mn}^j(\bm{x})\nonumber \\
        &-\frac{\hbar}{\pi c}\,\mathrm{p.v.} \sum_{m} \int_0^{\Lambda} dk\, \frac{k^{-1}}{E_{mn} + \hbar c k} \iint d\bm{x} d\bm{x}'\, \tau^{ij}\big(k\|\bm{x}'-\bm{x}\|\big) j_{P,nm}^i(\bm{x}') j_{P,mn}^j(\bm{x})\nonumber \\
        &-\frac{i}{\pi} \int_0^{\Lambda} dkk \iint d\bm{x}d\bm{x}'\, \sigma^{jk}\big(k\|\bm{x}'-\bm{x}\|\big) \expval{\big[m_P^k(\bm{x}'),p^j(\bm{x})\big]_-}_n.\label{eq3d152}
    \end{align}
Using that the curl of a gradient vanishes, through some index manipulations one can show that
\begin{equation}
   \left(-\delta^{ij} \partial^2 + \partial^i \partial^j\right)\varepsilon^{ipk}\partial'^pF(\bm{x},\bm{x}';k)\expval{\big[m_P^k(\bm{x}'),p^j(\bm{x})\big]_-}_n = k^2  \varepsilon^{jpk}\partial'^p F(\bm{x},\bm{x}';k)\expval{\big[m_P^k(\bm{x}'),p^j(\bm{x})\big]_-}_n
\end{equation}
and
\begin{equation}
    \sum_m E_{mn} p_{nm}^i(\bm{x}') p_{mn}^j(\bm{x}) = \frac{i\hbar}{2}\expval{\big[j_P^i(\bm{x}'), p^j(\bm{x})\big]_- - c \varepsilon^{ipk}\partial'^p\big[m_P^k(\bm{x}'), p^j(\bm{x})\big]_-}_n,
\end{equation}
we can collect the second and fourth lines together. Restoring the angular parts of the Fourier integrals, the second-order correction is
\begin{align}
    \Delta E_{n(2)}^{\Lambda} = \frac{i}{8\pi^2 c} \sum_{m} \int_{\|\bm{k}\|\leq \Lambda} d\bm{k}\, \|\bm{k}\|^{-1}\iint d\bm{x}d\bm{x}'\, \delta_T^{ij}(\bm{k})\, e^{i\bm{k}\cdot(\bm{x}'-\bm{x})} \expval{\big[j_P^i(\bm{x}'),p^j(\bm{x})\big]_- + c\varepsilon^{ipk}\partial'^p\big[m_P^k(\bm{x}'),p^j(\bm{x})\big]_-}_n\nonumber \\
    - \frac{\hbar}{4 \pi^2 c}\,\mathrm{p.v.} \sum_{m} \int_{\|\bm{k}\|\leq\Lambda} d\bm{k}\, \frac{\|\bm{k}\|^{-1}}{E_{mn} + \hbar \omega_{\bm{k}}} \iint d\bm{x}d\bm{x}'\, j_{P,nm}^i(\bm{x}') j_{P,mn}^j(\bm{x}) \delta_T^{ij}(\bm{k}) e^{i\bm{k}\cdot(\bm{x}'-\bm{x})}\nonumber \\
    - \frac{1}{\pi} \sum_{m} \int_0^{\Lambda} dk\, \iint d\bm{x} d\bm{x}'\, \tau^{ij}\big(k\|\bm{x}'-\bm{x}\|\big) p_{nm}^i(\bm{x}') p_{mn}^j(\bm{x}).
\end{align}
Working out the commutators in the first line, we find
\begin{align}
    \frac{1}{2}\expval{\big[j_P^i(\bm{x}'),p^j(\bm{x})\big]_- + c\varepsilon^{ipk}\partial'^p\big[m_P^k(\bm{x}'),p^j(\bm{x})\big]_-}_n = \frac{\hbar e}{mi} \bigg(\left[\frac{\partial}{\partial x^i} s^j(\bm{x};\bm{x}',\bm{R})\right]\rho_{nn}^e(\bm{x})\nonumber \\
    - \frac{1}{2} \int d\bm{y}\, \rho_{nn}^e(\bm{y})\left[\frac{\partial}{\partial y^k} s^i(\bm{x};\bm{y},\bm{R})\right]\left[\frac{\partial}{\partial y^k} s^j(\bm{x}';\bm{y},\bm{R})\right]\bigg),
\end{align}
and so, comparing to (\ref{cf1}), the second-order correction is (\ref{secondorder}).
\end{widetext}

\section{Energy shift in minimal coupling}\label{appendixd}
For comparison with the regularized energy shift (\ref{eq3d18}) computed with the multipolar Hamiltonian, we work out the same energy shift in minimal coupling. The minimal coupling Hamiltonian is given in the Schrödinger picture by equations (\ref{eq2a19}-\ref{eq28}). Split up the Hamiltonian into free and interaction terms
\begin{equation}
    H = H_0 + H_{\text{int}},
\end{equation}
where the free term is
\begin{equation}
    H_0 = H_B \otimes \mathbb{I}_F + \mathbb{I}_B \otimes H_F
    \label{H0mc}
\end{equation}
and the interaction term is
\begin{equation}
    H_{\text{int}} = - \frac{1}{c} \int d\bm{x}\, \bm{j}
    _P(\bm{x}) \cdot \bm{a}(\bm{x}) - \frac{1}{2c} \int d\bm{x}\, \bm{j}_D(\bm{x})\cdot\bm{a}(\bm{x}).
\end{equation}
The paramagnetic and diamagnetic current densities are defined in (\ref{minpara}) and (\ref{mindia}), respectively. As in Section \ref{section3}, we split up the interaction term 
\begin{equation}
    H_{\text{int}} = H_{\text{int}(1)} + H_{\text{int}(2)},
\end{equation}
where now
\begin{equation}
    H_{\text{int}(1)} = - \frac{1}{2c} \int d\bm{x}\, \bm{j}_D(\bm{x})\cdot\bm{a}(\bm{x})
\end{equation}
is already $\mathcal{O}(\alpha)$ and is treated at the first order, while
\begin{equation}
    H_{\text{int}(2)} = - \frac{1}{c} \int d\bm{x}\, \bm{j}_{P}(\bm{x})\cdot\bm{a}(\bm{x})
\end{equation}
is $\mathcal{O}(\sqrt{\alpha})$ and is treated at the second order. The first-order correction is
\begin{equation}
    \Delta E_{n(1)}^{\Lambda} = \frac{e^2}{2mc^2} \int d\bm{x}\, \bra{\Psi_{n}} \psi^{\dagger}(\bm{x}) \|\bm{a}(\bm{x})\|^2 \psi(\bm{x})\ket{\Psi_{n}},
\end{equation}
with the initial state (\ref{eq3c1}). In terms of the mode expansion (\ref{eqTempT}) for the vector potential,
\begin{equation}
    \Delta E_{n(1)}^{\Lambda} = \frac{\hbar^2 \alpha N_e}{\pi m} \int_0^{\Lambda} dk k = \frac{\alpha}{2\pi}\left(\frac{\hbar^2 N_e}{m}\right) \Lambda^2.\label{minfirstorder}
\end{equation}
\\
To obtain the second-order correction, begin with
\begin{equation}
    \bra{\Psi_{\delta}} H_{\text{int}(2)} \ket{\Psi_{n}} = - \frac{1}{c} \int d\bm{x}\, \bra{\Psi_{\delta}}\bm{j}_P(\bm{x})\cdot \bm{a}(\bm{x})\ket{\Psi_{n}}.
\end{equation}
Again using the mode expansion (\ref{eqTempT}), we have
\begin{equation}
    \abs{\bra{\Psi_{\delta}} H_{\text{int}(2)}\ket{\Psi_{n}}}^2 = \left(\frac{2\pi\hbar}{V\omega_{\bm{k}} }\right) \abs{\bm{e}_{I\bm{k}}\cdot \tilde{\bm{j}}_{P,nm}(\bm{k})}^2,
\end{equation} 
where $j_{P,mn}^j(\bm{x}) \equiv (j_P^j(\bm{x}))_{mn}$. Introducing the sums (\ref{eq3d4}) and taking the continuum limit, the regularized second-order correction is
\begin{widetext}
\begin{align}
    \Delta E_{n(2)}^{\Lambda} = - \frac{\hbar}{4\pi^2 c}\,\mathrm{p.v.} \sum_{m} \int_{\|\bm{k}\|\leq \Lambda} d\bm{k}\, \frac{\|\bm{k}\|^{-1}}{E_{mn} + \hbar \omega_{\bm{k}}}\sum_{I} \abs{\bm{e}_{I}(\bm{k}) \cdot \tilde{\bm{j}}_{P,mn}(\bm{k})}^2.\label{minsecondorder}
\end{align}
Collecting the first-order (\ref{minfirstorder}) and second-order (\ref{minsecondorder}) corrections together, the regularized energy shift at $\mathcal{O}(\alpha)$ in minimal coupling is exactly the result (\ref{eq3d18}).

\section{Proof of identity (\ref{eqTEMP14})}\label{appendixe}
We prove the identity (\ref{eqTEMP14}). To begin, consider the operator expression
\begin{equation}
    \frac{1}{H_F^R - E_n^R + \hbar \omega_{\bm{k}}} = \frac{1}{H_F^{(0)} - E_n^R + \hbar \omega_{\bm{k}}} - \frac{1}{H_F^R - E_n^R + \hbar \omega_{\bm{k}}} U^R \frac{1}{H_F^{(0)} - E_n^R + \hbar \omega_{\bm{k}}},
\end{equation}
from which follows
\begin{equation}
    \frac{1}{H_F^R - E_n^R + \hbar \omega_{\bm{k}}}\tilde{j}_P^j(\bm{k}) = \frac{1}{H_F^{(0)} - E_n^R + \hbar \omega_{\bm{k}}}\tilde{j}_P^j(\bm{k}) - \frac{1}{H_F^R - E_n^R + \hbar \omega_{\bm{k}}} U^R \frac{1}{H_F^{(0)} - E_n^R + \hbar \omega_{\bm{k}}} \tilde{j}_P^j(\bm{k}).
\end{equation}
To simplify, we will use the identity (\ref{intpicj}). Introducing a convergence factor ``$+i\delta$" (which will be removed below), we use an identity analogous to (\ref{rep}), namely
\begin{align}
    \frac{1}{H_F^{(0)} - E_n^R + \hbar \omega_{\bm{k}} + i \delta}\tilde{j}_P^j(\bm{k}) &= - i \int_0^{\infty} ds\, e^{is\left(\hbar \omega_{\bm{k}} - E_n^R + H_F^{(0)} + i\delta \right)} \tilde{j}_P^j(\bm{k})\nonumber \\
    &= - i \int_0^{\infty} ds\, e^{is\left(\hbar \omega_{\bm{k}} - E_n^R\ + i \delta \right)} e^{isH_F^{(0)}} \tilde{j}_P^j(\bm{k}) e^{-isH_F^{(0)}} e^{isH_{F}^{(0)}}\nonumber \\
    &= \tilde{j}_P^j(\bm{k}) \frac{1}{H_F^{(0)} - E_n^R + E_{\hbar\omega_{\bm{k}}} + \hbar \omega_{\bm{k}} + i \delta},
\end{align}
where we have dropped the $\|\bm{q}\|\ll mc$ terms as before. We can trivially take the limit $\delta \to 0^+$, since upon forming matrix elements there will never be poles in this denominator. Therefore we have
\begin{align}
    \frac{1}{H_F^R - E_n^R + \hbar \omega_{\bm{k}}}\tilde{j}_P^j(\bm{k}) &= \tilde{j}_P^j(\bm{k}) \frac{1}{H_F^{(0)} - E_n^R + E_{\hbar\omega_{\bm{k}}} + \hbar \omega_{\bm{k}}}\nonumber \\
    &- \frac{1}{H_F^R - E_n^R + \hbar \omega_{\bm{k}}} U^R \tilde{j}_P^j(\bm{k}) \frac{1}{H_F^{(0)} - E_n^R + E_{\hbar\omega_{\bm{k}}} + \hbar \omega_{\bm{k}}}\nonumber \\
    &= \tilde{j}_P^j(\bm{k}) \frac{1}{H_F^{(0)} - E_n^R + E_{\hbar\omega_{\bm{k}}} + \hbar \omega_{\bm{k}}}\nonumber \\
    &- \frac{1}{H_F^R - E_n^R + \hbar \omega_{\bm{k}}} \big[U^R, \tilde{j}_P^j(\bm{k})\big]_- \frac{1}{H_F^{(0)} - E_n^R + E_{\hbar\omega_{\bm{k}}} + \hbar \omega_{\bm{k}}}\nonumber \\
    &- \frac{1}{H_F^R - E_n^R + \hbar \omega_{\bm{k}}} \tilde{j}_P^j(\bm{k}) U^R \frac{1}{H_F^{(0)} - E_n^R + E_{\hbar\omega_{\bm{k}}} + \hbar \omega_{\bm{k}}},
\end{align}
and so bringing the last term to the left side
\begin{align}
    &\frac{1}{H_F^R - E_n^R + \hbar \omega_{\bm{k}}}\tilde{j}_P^j(\bm{k})\Big(H_F^R - E_n^R + E_{\hbar\omega_{\bm{k}}} + \hbar \omega_{\bm{k}}\Big) \frac{1}{H_F^{(0)} - E_n^R + E_{\hbar\omega_{\bm{k}}} + \hbar \omega_{\bm{k}}}\nonumber \\
    &= \tilde{j}_P^j(\bm{k}) \frac{1}{H_F^{(0)} - E_n^R + E_{\hbar\omega_{\bm{k}}} + \hbar \omega_{\bm{k}}} - \frac{1}{H_F^R - E_n^R + \hbar \omega_{\bm{k}}} \big[U^R, \tilde{j}_P^j(\bm{k})\big]_- \frac{1}{H_F^{(0)} - E_n^R + E_{\hbar\omega_{\bm{k}}} + \hbar \omega_{\bm{k}}}.
\end{align}
Therefore, cancelling the factor $(H_F^{(0)} - E_n^R + E_{\hbar\omega_{\bm{k}}} + \hbar \omega_{\bm{k}})^{-1}$ that features on both sides and multiplying the result by $(H_F^R - E_n^R + E_{\hbar\omega_{\bm{k}}} + \hbar \omega_{\bm{k}})^{-1}$ on the right, we end up with 
\begin{align}
    \frac{1}{H_F^R - E_n^R + \hbar \omega_{\bm{k}}}\tilde{j}_P^j(\bm{k}) &= \tilde{j}_P^j(\bm{k}) \frac{1}{H_F^{R} - E_n^R + E_{\hbar\omega_{\bm{k}}} + \hbar \omega_{\bm{k}}}\nonumber \\
    &- \frac{1}{H_F^R - E_n^R + \hbar \omega_{\bm{k}}} \big[U^R, \tilde{j}_P^j(\bm{k})\big]_- \frac{1}{H_F^R - E_n^R + E_{\hbar\omega_{\bm{k}}} + \hbar \omega_{\bm{k}}}.
\end{align}
Taking matrix elements in the states $\ket{\psi_{n,m}}$ and using that $H_F^R \ket{\psi_{n,m}} = E_{n,m}^R \ket{\psi_{n,m}}$, the desired result (\ref{eqTEMP14}) follows.
\end{widetext}

\section{Computations in Section \ref{section4}}
\subsection{Direct calculation of $\Delta E_n^R$}\label{appendixg}
To begin, we replace $E_{mn}^R$ in the denominator of (\ref{eq5T5}) by a ``reference" or average value $E_{mn}^R \to \bar{E}_n^R$ \cite{Bethe, MolQED}. Through a short calculation we have
\begin{align}
    \big\langle\big[U^R, \tilde{j}_P^j(\bm{x
    })\big]_-\big\rangle_{mn} = \frac{\hbar e}{m i} \expval{\psi^{\dagger}(\bm{x}) \partial^i \phi(\bm{x}) \psi(\bm{x})}_{nm}\nonumber \\
    + \frac{\hbar e}{m i} \expval{\psi^{\dagger}(\bm{x}) \partial^i \mathrm{V}(\bm{x}) \psi(\bm{x})}_{nm}.
\end{align}
Since we have replaced $E_{mn}^R$ in the denominator of (\ref{eq5T5}) by the reference value $\bar{E}_n^R$, we can evaluate the sum using the identity \cite{MolQED}
\begin{align}
    &\sum_m j_{P,nm}^i(\bm{x}) \big\langle\big[U^R, j_P^j(\bm{x}')\big]_{-} \big\rangle_{mn}\nonumber \\
    &= \frac{1}{2} \big\langle\big[\big[j_P^i(\bm{x}), U^R\big]_-, j_P^j(\bm{x}')\big]_-\big\rangle_{n}.\label{eq5T7}
\end{align}
\\
Evaluating this expression directly is cumbersome. It is easier to evaluate the complete expression
\begin{equation}
    \frac{1}{4\pi} \int d\Omega_{\bm{k}}\, \delta_T^{ij}(\bm{k}) \sum_m \tilde{j}_{P,nm}^i(-\bm{k}) \big\langle\big[U^R, \tilde{j}_P^j(\bm{k})\big]_-\big\rangle_{mn}.
\end{equation}
The angular integral yields $(2/3) \delta_{ij}$, and after a lengthy calculation we find
\begin{align}
    \frac{1}{4\pi} \int d\Omega_{\bm{k}}\, \delta_T^{ij}(\bm{k}) \sum_m \tilde{j}_{P,nm}^i(-\bm{k}) \big\langle\big[U^R, \tilde{j}_P^j(\bm{k})\big]_-\big\rangle_{mn}\nonumber \\
    = - \frac{\hbar^2 e^2}{3 m^2} \int d\bm{x}\, \expval{\psi^{\dagger}(\bm{x}) \nabla^2 \mathrm{V}(\bm{x})\psi(\bm{x})}_{n}. 
\end{align}
To arrive at this equality we have used the fact that $\psi(\bm{x})^2 = 0$ and $\psi^{\dagger}(\bm{x})^2 = 0$. Using the definition (\ref{eq2a2}) of the background ionic potential, 
\begin{align}
    \frac{1}{4\pi} \int d\Omega_{\bm{k}}\, \delta_T^{ij}(\bm{k}) \sum_m \tilde{j}_{P,nm}^i(-\bm{k}) \big\langle\big[U^R, \tilde{j}_P^j(\bm{k})\big]_-\big\rangle_{mn}\nonumber \\
    = - \frac{4\pi \hbar^2 e^2}{3 m^2} \sum_N q_N \expval{\rho^e(\bm{R} + \bm{d}_N)}_{n},
\end{align}
where $\bm{R} + \bm{d}_N$ is the location of the $N^{\text{th}}$ ion. Inserting this back into $\Delta E_{n}^R$, we find that our ``generalized Lamb shift" is given by
\begin{align}
    &\Delta E_{n}^R = - \frac{4 \alpha \hbar^3}{3 m^2 c}\sum_N q_N \expval{\rho^e(\bm{R}+\bm{d}_N)}_n\nonumber \\
    &\times \int_0^{\infty} dk\, \frac{1}{\hbar k / 2 m c + 1} \left(\frac{1}{\bar{E}_n^R + \hbar ck}\right).
\end{align}
Notice that this expression is \textit{finite}, and so the integral can be evaluated. The ``$\hbar k / 2m c$" term in the denominator accounts for spatial variations in the electromagnetic field. We can compute this integral using techniques from complex analysis. Define $\bar{\beta}_n^R \equiv \bar{E}_n^R/ 2 m c^2$ and consider the change of variables $x = \hbar k/2m c$. Then
\begin{align}
    \int_0^{\infty} dk\, \frac{1}{\hbar k / 2 m c + 1} \left(\frac{1}{\bar{E}_n^R + \hbar ck}\right)\nonumber \\
    = \left(\frac{1}{\hbar c}\right)^2 \int_0^{\infty}dx\, \frac{1}{(x + 1)(x + \bar{\beta}_n^R)}.
\end{align}
Consider the contour integral 
\begin{equation}
    \oint_{\mathscr{C}} dz \frac{\log(z)}{(z+1)(z + \bar{\beta}_n^R)},
\end{equation}
\\
where $\log(z)$ is the complex logarithm and $\mathscr{C}$ is the standard ``keyhole" contour that avoids the branch cut (at $\theta = 0$) of $\log(z)$ on the positive real axis. The contour integrals over the circular arcs vanish by application of the ``ML-estimate" method for complex integration \cite{Paliouras}. The other two integrals yield the identity
\begin{equation}
    \oint_{\mathscr{C}} dz \frac{\log(z)}{(z + 1)(z + \bar{\beta}_n)} = - 2\pi i \int_0^{\infty}dx\, \frac{1}{(x + 1)(x + \bar{\beta}_n)}.\label{temp102}
\end{equation}
Evaluating the contour integral on the left using the calculus of residues, we obtain
\begin{equation}
    \int_0^{\infty}dx\, \frac{1}{(x + 1)(x + \bar{\beta}_n^R)} = \left(\frac{1}{1 - \bar{E}_n^R / 2mc^2}\right) \ln(\frac{2mc^2}{\abs{\bar{E}_n^R}}),
\end{equation}
\\
and the result follows. 

\subsection{Calculation of the integral (\ref{contint})}\label{appendixf}
We calculate the integral
\begin{equation}
    \int_0^{\infty} dk \frac{k^{a+b} }{E_{mn}^R + \hbar ck }e^{- k / \Lambda'}
\end{equation}
using techniques from complex analysis. Introducing a change of variables $x \equiv \hbar k /2mc$, we have
\begin{widetext}
\begin{equation}
    \left(\frac{1}{\hbar c}\right)^2 \left(\frac{2mc}{\hbar}\right)^{a+b} \int_0^{\infty} dx\, \frac{{x}^{a+b}}{x + \beta_{mn}^R}e^{- x / \Lambda'} ,
    \label{intF3}
\end{equation}
\\
where $\beta_{mn}^R = E_{mn}^R / 2mc^2$. To evaluate this integral, consider the related contour integral
\begin{equation}
    \oint_{\mathscr{C}} dz \frac{z^{a+b} e^{-\abs{z}/\Lambda'}}{z + \beta_{mn}^R} \log(z),
\end{equation}
where $\mathscr{C}$ is the standard ``keyhole" contour. The integrals over the circular arcs $z = R e^{i\phi}$ and $z = \varepsilon e^{i\phi}$ vanish in the respective limits $R \to \infty$ and $\varepsilon \to 0^+$, by the ``ML estimate" for complex integration \cite{Paliouras}. Meanwhile, the remaining contour integrals, with the contours parametrized by $z = x \pm i \varepsilon$, can be combined together in the limit $\varepsilon \to 0^+$ and thereby yield the identity
\begin{equation}
    \int_0^{\infty} dx\, \frac{{x}^{a+b} e^{- x / \Lambda'}}{x + \beta_{mn}^R} = -\frac{1}{2\pi i}\oint_{\mathscr{C}} dz \frac{z^{a+b} e^{-\abs{z}/\Lambda'}}{z + \beta_{mn}^R} \log(z).
\end{equation}
We evaluate the integral on the right-hand-side using the calculus of residues: The pole is located at $z = - \beta_{mn}^R$ in the complex plane, and with the identity above we end up with
\begin{equation}
    \int_0^{\infty} dx\, \frac{{x}^{a+b} e^{- x / \Lambda'}}{x + \beta_{mn}^R} = - \big(-\beta_{mn}^R\big)^{a+b}e^{-\abs{\beta_{mn}^R}/\Lambda'} \ln(\abs{\beta_{mn}^R}),
\end{equation}
and so the integral (\ref{intF3}) is
\begin{equation}
    \left(\frac{1}{\hbar c}\right)^2 (-1)^{a+b} \left(\frac{E_{mn}^R}{\hbar c}\right)^{a+b} \ln(\frac{2mc^2}{\abs{E_{mn}^R}}),
    \label{intF2}
\end{equation}
where we have removed the regulator ($\Lambda'\to\infty$). With this result the desired expression (\ref{temp1000}) for the renormalized energy shift in its multipolar form follows.
\end{widetext}

\bibliographystyle{apsrev4-1}
\bibliography{AMOQEDManuscript.bib}

\end{document}